# La cadena de valor de la inteligencia artificial: una perspectiva crítica.
# The artificial intelligence value chain: A critical appraisal.
### [Spanish Version]

Pompeu Casanovas

*Profesor de Investigación* en el Instituto de Investigación en Inteligencia Artificial del Consejo Superior de Investigaciones Científicas (IIIA-CSIC, Barcelona, España)
*Investigador vinculado* al Instituto de Derecho y Tecnología de la Universidad Autónoma de Barcelona (IDT-UAB)
*Adjunct Professor*, Law School, La Trobe University, Melbourne, Australia
pompeu.casanovas@iiia.csic.es

**Resumen**. La cadena de valor de la inteligencia artificial es uno de los conceptos principales que vertebran la legislación europea sobre la materia, especialmente el *Reglamento de Inteligencia Artificial*. Se trata de un concepto de análisis económico que se ha convertido en jurídico —de gobernanza jurídica— a partir del uso continuado del mismo en documentos programáticos y en textos legales de la UE. El artículo (i) analiza su significación y su función en el marco de la estrategia de regulación establecido por los recientes programas de la UE —la *Brújula para la Competitividad*, el *Plan de Acción «Continente de IA»*, la *Estrategia de uso de la inteligencia artificial* y el *Ómnibus digital sobre la IA*—, (ii) señala sus limitaciones, y (iii) propone la construcción teórica de cadenas de valor que capten dimensiones intangibles que no son directamente monetizables, pero que tienen un gran impacto en el entorno social (como la lengua, la cultura, y, especialmente, los valores éticos y jurídicos). Compara también sucintamente tres marcos jurídicos distintos para la regulación de la IA (UE, Commonwealth y EE.UU). Al final, propone un marco para la construcción y análisis y análisis de la cadena ética y jurídica de valor como un medio para preservar los valores democráticos e incentivar la implementación digital del estado de derecho.

**Abstract.** The artificial intelligence value chain is one of the main concepts underpinning the European legislation on the subject, especially the *Artificial Intelligence Act*. It is an economic concept that has become a legal one—a concept of legal governance—due to its continued use in policy documents and legal texts. This article (i) analyses its significance and function within the framework of the regulatory strategy established by recent EU programs —the *Compass for Competitiveness*, the *«AI Continent» Action Plan, Apply AI Strategy,* and the *Digital Omnibus on AI* —, (ii) identifies its limitations, and (iii) advances the theoretical construction of value chains that capture intangible dimensions that are not directly monetizable (such as language, culture, and, especially, ethical and legal values) but have a significant impact on the social environment. It also briefly compares three different legal frameworks for the regulation of AI (EU, Commonwealth and USA). It proposes at the end a specific framework for the analysis of the ethical and legal AI value chain to preserve democratic values and foster the digital implementation of the rule of law.

**Secciones:**

# 1. Introducción

El presente artículo es fruto del panel sobre "Ética y responsabilidad en la aplicación de la inteligencia artificial a la lengua. La regulación legal", celebrado en Arequipa en el marco del X Congreso de la Lengua Española.[1] Como se verá, se sitúa un poco más allá de la presentación y del debate que tuvo lugar, pero va en su misma dirección y ahonda en su planteamiento.

Trataré primero el origen del concepto de *cadena de valor* (*value chain*, *CV* en adelante); en segundo lugar, el de cadena de valor *en* y *de* la IA (especialmente tal como se plantea en el Reglamento denominado «Ley Europea de IA»)[2]; y dejaré para el final la propuesta de una cadena de valor ética y jurídica, específica para la IA, y su relación con la lengua. Para mostrar desde el inicio a dónde me gustaría llegar, sostendré que el auténtico problema que aporta la introducción de la gestión de la información, el conocimiento y la tecnología misma mediante técnicas de IA generativa y de técnicas mixtas de IA simbólica y subsimbólica en el mercado, la sociedad y la administración *es el de la **reconstrucción de un espacio público en común, global y más justo** donde pueda operar la denominada* inteligencia 'híbrida'[3], 'coevolutiva'[4] *o simplemente* 'cointeligencia'[5] *entre humanos/as y máquinas* (*Human Machine Interaction*, *HMI*).

En este entorno, los instrumentos y conceptos de la teoría jurídica clásica siguen siendo esenciales, pero deben adaptarse mediante nuevos instrumentos de *gobernanza jurídica* que en cierto modo han modificado las fuentes mismas de las que partir en las distintas culturas jurídicas. Hoy veremos la cadena de valor de la IA; hay otros, como el de 'mejor regulación' (*better regulation*) o el de espacios protegidos de prueba y experimentación (*sandboxes*), que también deben tenerse en cuenta. Sin embargo, es crucial plantear que su éxito no puede suponerse: este plan es *propositivo*, no efectivo, y depende cada vez más de una geopolítica de relaciones inter y transnacionales muy tensa, con una escala de confrontación en aumento. Resulta relevante para ello volver a reflexionar sobre los instrumentos de los que disponemos para el desarrollo del estado de derecho —en el sentido de *rule of law* o *meta-rule of law* en la era digital, según nuestra formulación.[6] En la fase de uso generalizado de las técnicas de inteligencia artificial (IA), ello también implica al mismo tiempo una defensa estricta de los valores democráticos.

---

[1] Se trata de una versión extensa y ampliada de la comunicación presentada en el *X Congreso de la Lengua Española*, celebrado en Arequipa, Perú, del 13 al 18 de octubre de 2025. El panel 10.3 fue presidido por Julio del Valle Bayón, coordinado por Tomás de la Quadra Salcedo, y contó con la participación de Víctor Rodríguez Doncel, Ela Urriola Sanjur y José Luís Díaz Gómez, además de la mía propia. El objeto de análisis redactado por Tomás de la Quadra fue el siguiente: «Aprovechar los inmensos beneficios de la IA, evitando sus riesgos será el objetivo de su regulación. Experimentos con personas habituadas a ChatGPT demostrarían la disminución de capacidad de respuesta verbal o escrita cuando prescinden de la IA. La lengua y las palabras expresan emociones, sentimientos y valores que los algoritmos pueden recoger y remedar estadísticamente trasformados en datos, pero sólo desde una inteligencia sentiente (humana) podrán garantizarse respuestas o propuestas de actuación congruentes con la dimensión ética y moral de la acción humana, sin por ello prescindir, en absoluto, de la IA. Reflexión ética y regulación legal tendrán en cuenta lo que textos europeos denominan la "cadena de valor" de la IA, considerarán la factibilidad de "consciencia artificial", la capacidad de persuasión suprahumana, los rebeldes digitales y la intencionalidad y compromiso ético fundamentales para educar a usuarios y receptores de IA en el reconocimiento del origen humano de la obra artística o literaria.» Puede encontrarse el registro en vídeo de la sesión en el siguiente enlace: https://www.youtube.com/watch?v=zQbk3a4qRS8.

[2] Reglamento (UE) 2024/1689 del Parlamento Europeo y del Consejo, de 13 de junio de 2024. Cf. las referencias completas de los documentos, artículos y libros completos en la Sección final (10) de bibliografía.

[3] Cf. Akata et al. (2000).

[4] Pedreschi *et al*. (2025).

[5] Mollick (2024*).

[6] Cf. Casanovas (2015); Casanovas, de Koker y Hashmi (2022), Casanovas *et al*. (2022).

Este artículo —además de la Introducción (sección 1), las referencias bibliográficas y documentales (secciones 12-13), y los agradecimientos— consta de nueve secciones. La segunda sección (2) define tres marcos jurídicos de regulación según tres jurisdicciones con distintos instrumentos normativos —europea, británica (UK, más la Commonwealth), y estadounidense (USA). Australia y Canadá tienen modelos mixtos, entre UK y Europa, la primera; entre UK y USA, la segunda. La tercera sección (3) define el concepto de cadena de valor de la IA (CVIA), con relación a su origen en la CV genérica y en la estructura de empresa. Propone ya una definición alternativa de cadena de valor según su dimensión cultural, lingüística, ética y jurídica, y no únicamente económica. La sección 4 sitúa la cadena de valor de la IA (CVIA) respecto de conceptos afines —la cadena de valor de suministros (CVS) y la cadena global de valor (CGV). La sección 5 describe algunos obstáculos que hay que superar para redefinir la CVIA y propone el modo de superarlos mediate el establecimiento de siete supuestos básicos para efectuar esta redefinición. La sección 6 se centra en la regulación europea, i.e. el concepto de CVAI según es introducido en la denominada Ley Europea de Inteligencia Artificial, el Reglamento que entró en vigor el 1 de agosto de 2024 y será plenamente aplicable el 2 de agosto de 2026, con excepciones como la introducción en el mercado de prácticas prohibidas. Repasa también sus precedentes, estudios de impacto, y la legislación para reformarla de octubre y noviembre de 2025.  La sección 7 analiza el cambio de perspectiva que se ha producido en 2025 en la estrategia europea para el fomento de la innovación y la competencia. Este cambio es relevante para el concepto de CVAI que analizamos en el presente artículo. La sección 8 se centra en la *Propuesta de la Directiva sobre responsabilidad en materia de IA* que finalmente fue retirada a principios de 2025, y vuelve a plantear de manera más específica la necesidad de construir una cadena ética y jurídica de valor para la implementación de los sistemas de IA. La sección 9 expone algunas ideas relativas a la relación entre la cadena de valor de la IA y las lenguas naturales. Este era propiamente uno de los motivos del X Congreso, la relación de la IA con la lengua. Finalmente, la sección de conclusiones resume las tesis y señala algunas guías para el trabajo futuro.

## 2. Tres marcos jurídicos para la regulación de la IA

Desde el punto de vista de los gobiernos, estados, uniones y federaciones, se están desarrollando diversos marcos y estrategias para fomentar la innovación responsable en IA. Estos marcos suelen adoptar un enfoque basado en el riesgo, imponiendo requisitos más estrictos a los sistemas de IA con el potencial de causar un mayor daño social. Dejando aparte la vía de regulación china[7], hay por lo menos tres estrategias diferentes que corresponden al distinto tipo de cultura jurídica de Europa, los países de la Commonwealth y Estados Unidos:[8]

1. El enfoque *holístico o integral* de la UE, mediante la Ley de IA, entre otras, crea un marco jurídico aplicable a todos los sectores. Utiliza un modelo basado en la pirámide de riesgos, prohibiendo ciertas aplicaciones de "riesgo inaceptable" (como la construcción de perfiles, *social scoring*) e imponiendo requisitos estrictos a los sistemas de "alto riesgo" (como los utilizados en dispositivos médicos).

---

[7] Cf. el programa de industrialización basado en una fuerte inversión en IA *Made in China 2025*, CSET (2022, marzo).
[8] El lector puede encontrar una somera descripción y análisis de la normativa emergente sobre IA (hasta 2020) en Australia, Canadá, Alemania, UK, Japón, Singapur, USA, más algunos estándares (e.g. IEC-ISOs) en el estudio complementario de la evaluación de impacto del Reglamento de IA encargado por la Comisión. Cf. Renda *et al*. (2020, pp. 75-101).

2. El enfoque *basado en principios*. Esta estrategia «pro-innovación», adoptada por el Reino Unido, se basa en un conjunto de cinco principios generales adoptados desde el *White Paper* en IA (2023) —*seguridad, transparencia, equidad, rendición de cuentas e impugnación*)—[9] para ser aplicados por los reguladores en cada sector público.[10] En lugar de crear una nueva ley específica, aprovecha los organismos reguladores existentes (como los de finanzas o salud) para hacer cumplir estos principios. Otros países de la Commonwealth —especialmente Australia[11]— han adoptado una estrategia mixta, partiendo de principios y de la legislación ya existente, pero acercándose al modelo europeo en relación a las situaciones de riesgo.[12]
3. El enfoque *fragmentado o sectorial*. Éste es el modelo aplicado en Estados Unidos, que carece de una ley federal única e integral. En cambio, la regulación es un «mosaico de regulaciones», *a patchwork kilt* (una «falda escocesa», como lo denomina la UE) de leyes existentes, estándares del *National Institute of Standardization and Technology*,

---

[9] Los principios son: 1. Protección, seguridad y robustez (*Safety, security and robustness*); 2. Transparencia y explicabilidad adecuadas (*Appropriate transparency and explainability*); 3. Justicia y equidad (*Fairness*); 4. Responsabilidad y gobernanza (*Accountability and governance*); 5. Impugnación y reparación (*Contestability and redress*). Cf. UK Department for Science, Innovation and Technology (2023).

[10] Cf. UK Department for Science, Innovation and Technology (febrero, 2024).

[11] Hay una tensión en Australia entre la más tradicional autorregulación o «regulación voluntaria» basada en principios y la propuesta de política legislativa basada en diez «barreras de protección obligatorias» (*mandatory guardrails*) para la IA de alto riesgo. Esta política se acerca a la política establecida por la *AI Act* europea. Las diez medidas son las siguientes: 1. Establecer un proceso de rendición de cuentas que incluya gobernanza, capacidad interna y una estrategia para el cumplimiento normativo. 2. Establecer un proceso de gestión y mitigación de riesgos. 3. Implementar medidas de gobernanza de datos para gestionar la calidad y la procedencia de los datos. 4. Probar los modelos y sistemas de IA para evaluar su rendimiento y supervisar el sistema una vez implementado. 5. Permitir el control o la intervención humana en un sistema de IA para lograr una supervisión humana efectiva. 6. Informar a los usuarios finales sobre las decisiones basadas en IA, las interacciones con la IA y el contenido generado por IA. 7. Establecer procesos para que las personas afectadas por los sistemas de IA puedan cuestionar su uso o sus resultados. 8. Ser transparentes con otras organizaciones de la cadena de suministro de IA sobre los datos, los modelos y los sistemas para ayudarlas a abordar los riesgos de manera eficaz. 9. Mantener registros que permitan a terceros evaluar el cumplimiento de las salvaguardias. 10. Realizar evaluaciones de conformidad para demostrar y certificar el cumplimiento de las salvaguardias. CF. AUS Gov. (2024, Sept.). Con posterioridad, los principios han sido sintetizados en seis prácticas para la industria, y su obligatoriedad puesta en cuestión por las políticas de incentivos industriales. Cf. AUS Gov. (2025, 5 de agosto); AUS Gov. (2025a, octubre) y AUS Gov. (2025b, octubre).

[12] El modelo de Canadá se acerca más al de regulación sectorial mediante estándares de los EUA, y se ha visto afectado por los cambios de gobierno. Había sido prevista una legislación específica para 2025 —*Artificial Intelligence and Data Act (AIDA): As part of Bill C-27*— que no ha llegado a aprobarse y a entrar en vigor. El reciente Estándar CAN/DGSI 101:2025 para pequeñas y medianas empresas (hasta 500 trabajadores) proporciona un marco de referencia y un proceso específico para organizaciones que utilizan IA con aprendizaje automático para la toma de decisiones. Tiene también desde 2023 un código de «regulación voluntaria» con los siguientes principios: 1. *Responsabilidad* (*accountability*) Las organizaciones comprenden su función con respecto a los sistemas que desarrollan o gestionan, implementan sistemas de gestión de riesgos adecuados y comparten información con otras organizaciones según sea necesario para evitar deficiencias. 2. *Seguridad* (*safety*): Los sistemas se someten a evaluaciones de riesgos y se implementan las medidas de mitigación necesarias para garantizar un funcionamiento seguro antes de su puesta en marcha. 3. *Equidad* e *imparcialidad* (*fairness and equity*): Se evalúan y abordan los posibles impactos en materia de equidad e imparcialidad en las diferentes fases de desarrollo y puesta en marcha de los sistemas. 4. *Transparencia* (*transparency*): Se publica información suficiente para que los consumidores puedan tomar decisiones informadas y para que los expertos evalúen si los riesgos se han abordado adecuadamente. 5. *Supervisión y control humanos* (*human oversight and monitoring*): Se supervisa el uso del sistema tras su puesta en marcha y se implementan las actualizaciones necesarias para abordar cualquier riesgo que se materialice. 6. *Validez y robustez* (*validity and robustness*): Los sistemas funcionan según lo previsto, son seguros frente a ciberataques y se comprende su comportamiento en respuesta a la variedad de tareas o situaciones a las que probablemente estén expuestos. Cf. CAN (2023) y CAN (2025).

NIST, órdenes presidenciales y órdenes ejecutivas sectoriales.[13] Aunque puede generar inconsistencias y lagunas de regulación, este enfoque se caracteriza por su adaptabilidad y flexibilidad en la aplicación. Como veremos enseguida, se está produciendo actualmente una centralización presidencial de este enfoque.

El problema es que la cooperación entre estas estrategias de regulación no puede darse por sentada, sino que produce escenarios cuyo marco regulatorio se halla también fragmentado y en tensión mutua. Hay no solamente disgregación y desagregación, sino competición y falta de acuerdo entre los actores y reguladores (sean éstos corporaciones, estados nacionales u organizaciones empresariales). Bajo el mandato de Donald Trump, se han sucedido las Órdenes Presidenciales ejecutivas para asegurar el predominio norteamericano en el desarrollo de la IA.[14] Así, la denominada *Misión Génesis* (*Genesis Mission*) de 24 de noviembre de 2025[15], y la Orden Presidencial de 11 de diciembre de 2025 sobre la construcción de un único marco nacional para el desarrollo de las políticas de IA bajo el control directo de Presidencia. Ésta última Orden establece claramente que:

> Para ganar, las empresas estadounidenses de IA deben tener la libertad de innovar sin una regulación engorrosa. Sin embargo, la excesiva regulación estatal frustra este imperativo. En primer lugar, la regulación estatal, por definición, crea un mosaico de 50 regímenes regulatorios diferentes que dificulta el cumplimiento, especialmente para las empresas emergentes. En segundo lugar, las leyes estatales son cada vez más responsables de exigir a las entidades que incorporen sesgos ideológicos en los modelos. Por ejemplo, una nueva ley de Colorado que prohíbe la discriminación algorítmica puede incluso obligar a los modelos de IA a producir resultados falsos para evitar un trato o impacto diferenciado en los grupos protegidos. En tercer lugar, las leyes estatales a veces regulan de forma inadmisible más allá de las fronteras estatales, lo que afecta al comercio interestatal.[16]

Se trata de una desregulación orquestada y controlada en beneficio de las corporaciones y de la posición de dominio de Estados Unidos en el mercado global. Podríamos representarla como un corte del nudo gordiano de la regulación estatal de la cadena de valor para subordinarla a un único objetivo: «*A carefully crafted national framework can ensure that the United States wins the AI race, as we must. [...] It is the policy of the United States to sustain and enhance the United States' global AI dominance through a minimally burdensome national policy framework for AI*». Esto no es propiamente nuevo. La desregulación políticamente centralizada había sido anticipada desde el principio por los Memoranda[17] y primeras Órdenes Ejecutivas del presidente.[18] *Remove Red Tape and Onerous Regulation* es el principal mensaje del *America's AI Action Plan*.[19] El segundo es la descalificación como «amenaza existencial» de las políticas DEI (*diversity, equity and inclusión*) y de la «ideología *WOKE*».[20]

---

[13] Cf. Szczepański (2024), siguiendo la expresión utilizada en el CSIS por Bill Whyman (2023).
[14] Pueden consultarse las Órdenes Ejecutivas sucesivas en https://www.whitehouse.gov/presidential-actions/executive-orders/ Cf. especialmente, US Executive Order (2025, noviembre 24): «Hoy, Estados Unidos se encuentra en una carrera por el dominio tecnológico global en el desarrollo de la inteligencia artificial (IA), una importante frontera del descubrimiento científico y el crecimiento económico. Con ese fin, mi Administración ha tomado diversas medidas para ganar esa carrera, incluyendo la emisión de múltiples Órdenes Ejecutivas y la implementación del Plan de Acción de IA de Estados Unidos, que reconoce la necesidad de invertir en ciencia basada en IA para acelerar el avance científico. En este momento crucial, los desafíos a los que nos enfrentamos requieren un esfuerzo nacional histórico, comparable en urgencia y ambición al Proyecto Manhattan, que fue decisivo para nuestra victoria en la Segunda Guerra Mundial y una base fundamental para la fundación del Departamento de Energía) y sus laboratorios nacionales».
[15] *Ibid*. US EO 2025, noviembre 24: «Esta orden lanza la "Misión Génesis", un esfuerzo nacional dedicado y coordinado para impulsar una nueva era de innovación y descubrimiento acelerados por la IA que pueda resolver los problemas más desafiantes de este siglo. La Misión Génesis construirá una plataforma integrada de IA para aprovechar los conjuntos de datos científicos federales —la mayor colección de este tipo de datos del mundo, desarrollada a lo largo de décadas de inversión federal— para entrenar modelos de base científica y crear agentes de IA que prueben nuevas hipótesis, automaticen los flujos de trabajo de investigación y aceleren los avances científicos».
[16] Cf. US EO 2025, diciembre 11.

La UE, por su parte, ha reaccionado contra esta explícita política de confrontación y predominio mediante dos documentos programáticos en enero y abril de 2025: (i) una «Brújula para la Competitividad de la UE», que anuncia una nueva ley de materiales avanzados (*Materials Advanced Act*) para ser promulgada en 2026 y «apoyar todo el ciclo de vida, desde la investigación y la innovación hasta la creación de empresas emergentes, pasando por la fabricación y la implementación»;[21] (ii) y «un «Plan de Acción» que intenta subrayar los puntos fuertes europeos y que anuncia una futura *Ley de desarrollo de la nube y la inteligencia artificial*.[22] Así:

> La UE debe mantener su propio enfoque distintivo de la IA aprovechando sus puntos fuertes y lo que mejor hace. Ello incluye: en primer lugar, un gran mercado único con un único conjunto de normas de seguridad en toda la Unión, incluido el Reglamento de Inteligencia Artificial recientemente adoptado, que garantice que la IA sea fiable y esté en consonancia con los valores de la Unión; en segundo lugar, un aprovechamiento máximo de su investigación científica de alta calidad, una importante reserva de científicos y profesionales cualificados; en tercer lugar, una próspera escena de empresas emergentes y en expansión, conocimientos industriales y experiencia, y, por último, pero no por ello menos importante, una base sólida de potencia computacional de categoría mundial con espacios de datos accesibles para todos.[23]

---

[17] Cf. United States Presidential Memorandum (2025, 22 marzo) sobre la actuación jurídica y judicial de 22 de marzo 2025 para «prevenir los abusos» de los despachos de abogados y de la propia judicatura.

[18] *Vid*. por ejemplo, la US EO 2025, enero 23, donde se anunciaba la abrogación de todas las órdenes presidenciales y legislación anteriores que pusieran obstáculos al desarrollo de la IA: «La política de los Estados Unidos consiste en sostener y mejorar su dominio global en materia de inteligencia artificial para promover el florecimiento humano, la competitividad económica y la seguridad nacional». Cf. asimismo la US EO 2025, de 6 de abril, sobre la eliminación de barreras regulatorias para la libre competencia: «Las regulaciones federales no deberían predeterminar quiénes son los ganadores y perdedores económicos. Sin embargo, algunas regulaciones operan para excluir a nuevos participantes del mercado. Las regulaciones que reducen la competencia, el emprendimiento y la innovación, así como los beneficios que generan para los consumidores estadounidenses, deben ser eliminadas. La presente Orden inicia el proceso de eliminación de regulaciones anticompetitivas y la revitalización de la economía estadounidense».

[19] The White House (2025, 10 julio, p. 3).

[20] Cf. US Executive Order 2025 de 23 de julio: «La inteligencia artificial (IA) desempeñará un papel fundamental en la forma en que los estadounidenses de todas las edades adquieren nuevas habilidades, consumen información y desenvuelven su vida diaria. Los estadounidenses requerirán resultados fiables de la IA, pero cuando se incorporan sesgos ideológicos o agendas sociales en los modelos de IA, pueden distorsionar la calidad y la precisión de los resultados. Una de las ideologías más extendidas y destructivas es la denominada "diversidad, equidad e inclusión" (DEI). En el contexto de la IA, la DEI incluye la supresión o distorsión de información factual sobre raza o sexo; la manipulación de la representación racial o sexual en los resultados de los modelos; la incorporación de conceptos como la teoría crítica de la raza, el transgenerismo, el sesgo inconsciente, la interseccionalidad y el racismo sistémico; y la discriminación por motivos de raza o sexo. La DEI desplaza el compromiso con la verdad en favor de resultados preferentes y, como ilustra la historia reciente, representa una amenaza existencial para una IA fiable».

[21] Cf. EU Commission (2024, febrero, p.1): «Se entiende por materiales avanzados los materiales diseñados racionalmente para tener i) propiedades nuevas o mejoradas, o ii) características estructurales específicas o mejoradas con el objetivo de lograr un rendimiento funcional específico o mejorado. Esto incluye tanto los materiales fabricados emergentes nuevos (materiales de alta tecnología) como los materiales fabricados a partir de materiales tradicionales (materiales de baja tecnología)». Vid. asimismo, https://research-and-innovation.ec.europa.eu/news/all-research-and-innovation-news/commission-seeks-feedback-future-advanced-materials-act-2025-10-21_en .

[22] EU Cloud and AI Development Act. Cf. https://www.eu-cloud-ai-act.com/ .

[23] EU Commission (2025, abril, p.1). Plan de Acción «Continente de IA». Cito según la versión española oficial.

Sin embargo, quizás movida por la urgente necesidad de reacción, la UE ha anticipado objetivos difícilmente alcanzables —como el de la inteligencia artificial general[24]— de los que hablaremos luego y, después de los Informes de Mario Draghi y Enrico Letta[25], apuesta por una simplificación de los requisitos administrativos —creados por ella misma, por cierto— para la integración de la IA y la innovación en los modelos productivos.[26] Constituirá, según sus propias palabras, un «28.º régimen jurídico» añadido al de los veintisiete países de la UE como nuevo marco de regulación.[27] Sirva como muestra de la ambición del programa y también de la imprecisión normativa que puede detectarse en la equiparación de jurisdicciones distintas.

La comparación de la regulación europea de la IA con la propia de Estados Unidos y de la *Commonwealth* es ilustrativa. La regulación europea tiene un sesgo de naturaleza estatal con una doble vertiente. Está, por un lado, la tensión competitiva con la soberanía de los estados miembros. Y, por otro, la que deriva internamente de los modos de regulación basados en normas ejecutivas —como los Reglamentos—, que deben compatibilizarse con otras categorías de normas más horizontales (básicamente estándares, protocolos y códigos éticos). En Europa, la adopción de medidas de gobernanza jurídica basadas en la coregulación y la adopción de estándares ha sido mucho más lenta y confusa. En la práctica, esto la convierte en más compleja técnicamente.[28] Debe

---

[24] «En los dos últimos años, *los modelos de IA se han vuelto cada vez más complejos, pasando del procesamiento de textos al razonamiento, las capacidades multimodales y el comportamiento de agente*. Esta tendencia continuará, *y se espera que la próxima generación de modelos fronterizos de IA desencadene un salto en las capacidades, hacia la inteligencia artificial general capaz de realizar tareas muy complejas y diversas, igualando las capacidades humanas*». *Ibid.* p. 8 [El primer subrayado es del texto; el segundo es añadido].

[25] Draghi (2024, septiembre; 2025, septiembre); Letta (abril, 2024). Letta (*ibid.*, p. 40) insiste especialmente en la noción de cadena de valor para evaluar los efectos positivos de los denominados «Proyectos Importantes de Interés Común Europeo» (IPCEI, en inglés): «Para muchas de las ambiciones de la política industrial de la UE, el modelo IPCEI podría servir de modelo, ya que incorpora un enfoque comparativamente europeo para la formulación de políticas públicas. En primer lugar, los IPCEI deben contribuir a los objetivos de la UE. En segundo lugar, estos proyectos deben demostrar su capacidad para superar importantes fallos del mercado o del sistema, o para abordar retos sociales clave. En tercer lugar, los IPCEI deben involucrar a empresas ubicadas en varios Estados miembros y brindar a todos ellos la oportunidad de participar. *Por último, estos proyectos deben generar efectos indirectos positivos más allá de los países participantes, adoptando un enfoque de cadena de valor* [énfasis añadido]. Sobre los IPCEI, cf. https://competition-policy.ec.europa.eu/state-aid/ipcei_en .

[26] Cf. EU Commission (2025, enero, p.3). «La Brújula persigue dos objetivos generales. En primer lugar, identificar los cambios políticos necesarios para que Europa avance a un ritmo superior. En algunos ámbitos, será necesario modernizar las políticas existentes; en otros, se requiere un cambio radical para adaptarse a las nuevas realidades. El segundo objetivo es desarrollar nuevas formas de colaboración para aumentar la velocidad y la calidad de la toma de decisiones, simplificar nuestros marcos y normas, y superar la fragmentación.»

[27] EU Commission (2025, enero, p. 5). «Permitir que las empresas innovadoras se beneficien de un conjunto único y armonizado de normas a nivel de la UE dondequiera que inviertan y operen en el Mercado Único, en lugar de enfrentarse a 27 regímenes jurídicos distintos, representaría un verdadero punto de inflexión. Por ello, la Comisión propondrá un 28.º régimen jurídico, que simplificará las normas aplicables y reducirá el coste del fracaso, incluyendo todos los aspectos pertinentes del derecho de sociedades, la insolvencia, el derecho laboral y el derecho fiscal.»

[28] Véase, en este sentido, Micklitz (2023) y Micklitz (2024). En la proposición de la nueva ley *Ómnibus* de 19 de noviembre de 2025, Considerando 2, se reconoce expresamente el retraso en la elaboración y aplicación de estándares: «La experiencia adquirida en la aplicación de las partes del Reglamento (UE) 2024/1689 que ya han entrado en vigor puede servir de base para la aplicación de las partes que aún no lo han hecho. En este contexto, el retraso en la elaboración de estándares, que deberían proporcionar soluciones técnicas a los proveedores de sistemas de IA de alto riesgo para garantizar el cumplimiento de sus obligaciones en virtud de dicho Reglamento, y el retraso en el establecimiento de los marcos de gobernanza y evaluación de la conformidad a nivel nacional resultan en una carga de cumplimiento superior a la prevista. Además, las consultas con las partes interesadas han puesto de manifiesto la necesidad de medidas adicionales que

compatibilizarse además con una autorregulación aparente, i.e. con la obligación de proveedores y desarrolladores de probar el cumplimiento de los requisitos impuestos por las provisiones legislativas (especialmente en la Ley de Inteligencia Artificial).

Dos iniciativas muy recientes más intentan paliar este déficit regulatorio para incentivar la implementación de la IA y desarrollar el programa político descrito. Hay que añadirlas a la «Brújula» y al Plan de Acción («Continente») ya mencionados. Se trata de: (iii) la *Estrategia de uso de la IA* (en octubre de 2025), y (iv) la reforma de artículos relevantes de la Ley de la IA mediante la propuesta de Reglamento *Omnibus Digital sobre la IA* (de 19 de noviembre de 2025). Volveremos brevemente sobre estas nuevas iniciativas más adelante (*vid. supra*, sección 7), porque intentan facilitar el cumplimiento de los participantes en el ciclo de vida e implementación de los sistemas de IA sin reducir por ello los requerimientos para su aprobación y de la pirámide de riesgos. Vayamos ahora a la cadena de valor.

### 3. La cadena de valor de la IA (CVIA)

La denominada *cadena de valor de la inteligencia artificial* es uno de los conceptos básicos del Reglamento de Inteligencia Artificial europeo, no modificada (antes, al contrario) por la propuesta *Omnibus*. De hecho, se trata de un concepto que también se halla presente en la estrategia de regulación de la Commonwealth y de Estados Unidos, sin definirlo explícitamente como la base o eje central de su desarrollo jurídico. La razón estriba en que el concepto mismo de cadena de valor se considera hoy en día un supuesto del desarrollo de la ventaja competitiva de las empresas y corporaciones en el mercado.

El libro que primero definió la *cadena genérica de valor* llevaba este título, *Competitive Advantage*[29], y marcó época en el ámbito de la gestión y estrategia corporativa. Michael E. Porter la definía como un proceso estructurado de desagregación de la empresa en las actividades que subyacen a la ventaja competitiva e identifica las relaciones entre ellas.[30] Así, la cadena de valor "desagrega una empresa en sus actividades estratégicamente relevantes para comprender el comportamiento de los costes y las potenciales fuentes de diferenciación existentes".[31] La Fig. 1 muestra el esquema general de la cadena de valor. Constituye un esquema que ha sido reproducido en múltiples ocasiones y que ayuda a distinguir los distintos niveles y la relación de las *actividades de soporte* —infraestructura de la empresa, gestión de recursos humanos, desarrollo de tecnologías, abastecimiento— y las *actividades primarias* —logística de entrada, operaciones, logística de salida, marketing y ventas, servicios. ***El concepto se refiere claramente a la organización estratégica de empresas y corporaciones en la competencia por los beneficios en el mercado.***

---

faciliten y aclaren la aplicación y el cumplimiento, sin reducir el nivel de protección de la salud, la seguridad y los derechos fundamentales frente a los riesgos relacionados con la IA que las normas del Reglamento (UE) 2024/1689 pretenden alcanzar.» Cf. EU Commission (noviembre 2025) (*Digital Omnibus on AI*).
[29] Porter (1985).
[30] Ibid. p. 27.
[31] Ibid. p. 33.

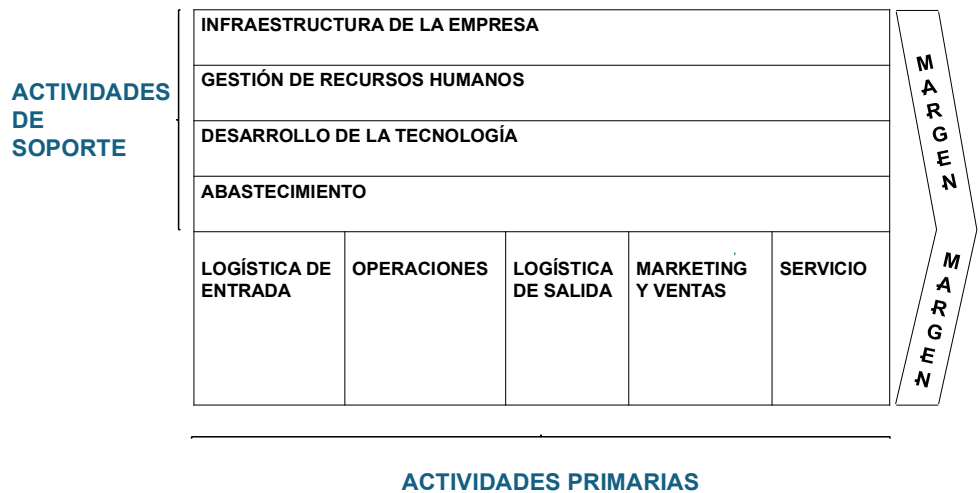

**Fig. 1**. Cadena genérica de valor. Fuente: Porter (1985).[32]

Siguiendo este esquema, la cadena de valor de la IA se compondría de una serie de *actividades primarias y de soporte*. Las actividades de soporte comprenderían funciones esenciales: (i) la investigación y la innovación de algoritmos, (ii) la gobernanza de datos y la supervisión ética, (iii) el cumplimiento normativo y jurídico, (iv) y la gestión de riesgos. Estamos refiriéndonos a actividades cruciales para mitigar los riesgos y garantizar el funcionamiento responsable y sostenible de toda la cadena de valor económica.

Las actividades primarias consistirían en: (i) *Adquisición y etiquetado de datos*: El paso inicial consiste en recopilar datos sin procesar de diversas fuentes y prepararlos para el entrenamiento del modelo. Este proceso es fundamental, ya que la calidad de los datos influye directamente en el rendimiento y el perfil ético del modelo. (ii) *Diseño de la arquitectura del modelo*: El plano conceptual del modelo de IA. (iii) *Entrenamiento y optimización del modelo*: El proceso computacionalmente intensivo de entrenamiento de un algoritmo con grandes conjuntos de datos. (iv) *Infraestructura y gestión computacional*: La provisión y gestión de los recursos de hardware y software necesarios para el entrenamiento y la implementación. (v) *Implementación e integración*: El proceso de hacer operativo el modelo de IA, a menudo mediante interfaces de programación de aplicaciones (API), e integrarlo en un producto o servicio existente. (vi) *Monitorización y aprendizaje*: El proceso continuo de observar el rendimiento del modelo, detectar anomalías y actualizarlo para incorporar nuevos datos y mejorar la precisión. (vii) *Conversión en producto e integración en el mercado* del modelo: Los pasos finales para convertir el modelo de IA en un producto (sistema) viable para el mercado. (viii) *Atención al cliente*: Brindar soporte a los usuarios finales.

Ésta es una síntesis de las posibilidades existentes.[33] Obsérvese el sesgo en las actividades de soporte, las cuales incluirían todos los aspectos normativos (éticos y jurídicos), gestionados por los departamentos de recursos humanos o los de cumplimiento normativo (*compliance*). La ISO

---

[32] Cf. el diagrama original de la cadena genérica de valor en Porter (1985, p. 37).
[33] He consultado GEMINI después de establecer el estado del arte de la cadena de valor. Véase el sesgo observado, que debe ponerse en relación con los estándares internacionales ISO, con los trabajos de la OECD, NIST, y con la tendencia general de los analistas económicos de situar la dimensión ética y jurídica exclusivamente dentro de las actividades de soporte.

9001: 2015 incorporó el análisis de Porter al control de calidad de las organizaciones.[34] La ISO 37301: 2021 ha reemplazado a la ISO 19600: 2014 y establece los requisitos para implementar un sistema de gestión basado en el cumplimiento normativo (*compliance*).[35] Los analistas corporativos de datos han empezado a utilizar niveles y patrones similares para mostrar las ventajas y desventajas de las empresas con relación a las fluctuaciones del mercado de los grandes modelos de lenguaje (*Large Language Models*, LLM) y modelos fundacionales. La aparición de Deep Seek en enero de 2025 conllevó una mayor atención al concepto, para marcar las pérdidas y ganancias en la cadena de valor.[36] Hay quién ha empezado a establecer también el valor en la cadena de suministros de los grandes modelos de lenguaje (*LLM Supply Chain*)[37] y, a raíz de la Ley de IA, la «cadena de valor de un sistema de IA de uso general» (*the value chain of general-purpose AI*),[38] i.e. «sistema de IA basado en un *modelo de IA de uso general* (*general-purpose AI model*) y que puede servir para diversos fines, tanto para su uso directo como para su integración en otros sistemas de IA».[39]

Es de notar que este último término, antes inexistente, no había sido utilizado en el ámbito de las ciencias de la computación o del diseño. Prosigue a grandes rasgos la distinción usual entre IA 'fuerte' y 'débil'.[40] El Reglamento define en su artículo 3 categorías que no guardan una

---

[34] Esto es más que una valoración. Corresponde a la descripción de la norma realizada por la propia organización de estandarización. Cf. https://co.isotools.us/iso-90012015-cadena-valor-porter/ .

[35] Cf. https://www.iso.org/obp/ui#iso:std:iso:37301:ed-1:v1:es . Debe ponerse en relación con la ISO/IEC TR 5469:2024, que proporciona directrices sobre el uso de la Inteligencia Artificial (IA) en la seguridad funcional. Comprende propiedades, factores de riesgo y métodos para sistemas relacionados con la seguridad. El estándar detalla cómo integrar la IA en las funciones de seguridad y utilizar métodos no basados en IA. Cf. https://www.iso.org/standard/81283.html.

[36] Cf. Wegner (2025). IOT Analytics realizó en febrero de 2025 la comparación entre las pérdidas y ganancias que supuso la aparición de Deep Seek en el mercado para las diversas empresas proveedoras. El análisis distinguía entre «ganadores» y «perdedores». Se proponían diversas capas (*tiers*) de complejidad: (i) *Usuarios finales*, incluyendo consumidores y empresas que utilizan una aplicación de IA generativa; (ii) *Aplicaciones* GenAI: Proveedores de *software* que incluyen funciones GenAI en sus productos u ofrecen software GenAI independiente (empresas de software empresarial como Salesforce, centrada en la IA de agentes, y *startups* especializadas en aplicaciones GenAI como Perplexity o Lovable; (iii) *Beneficiarios de nivel 1*, i.e. proveedores de modelos base (p. ej., OpenAI o Anthropic), plataformas de gestión de modelos (p. ej., AWS Sagemaker, Google Vertex o Microsoft Azure AI), herramientas de gestión de datos (p. ej., MongoDB o Snowflake), computación en la nube y operaciones de centros de datos (p. ej., Azure, AWS, Equinix o Digital Realty), consultores de IA y servicios de integración (p. ej., Accenture o Capgemini) y computación en el borde (p. ej., Advantech o HPE); (iv) *Beneficiarios de nivel 2*: Aquellos cuyos productos y servicios respaldan regularmente los servicios de nivel 1, incluyendo proveedores de chips (p. ej., NVIDIA o AMD), equipos de red y servidores (p. ej., Arista Networks, Huawei o Belden) y tecnologías de refrigeración de servidores (p. ej., Vertiv o Schneider Electric). (v) *Beneficiarios de nivel 3*: Aquellas empresas cuyos productos y servicios respaldan regularmente los servicios de nivel 2, como proveedores de software de automatización de diseño electrónico para diseño de chips (p. ej., Cadence o Synopsis), fabricación de semiconductores (p. ej., TSMC), intercambiadores de calor para tecnologías de refrigeración y tecnología de redes eléctricas (p. ej., Siemens Energy o ABB); (vi*). Beneficiarios de nivel 4 y superiores*: Empresas que siguen respaldando el nivel superior, como sistemas de litografía (nivel 4) necesarios para máquinas de fabricación de semiconductores (p. ej., AMSL) o empresas que proporcionan a estos proveedores (nivel 5) óptica litográfica (p. ej., Zeiss).

[37] Cf. Wang *et al*. (2025). Véase la definición en *infra*, nota 47.

[38] Cf, Küspert , Moës y Dunlop (2023).

[39] Ley de IA (EU Co, 2024), Art. 3.66.

[40] El término se encuentra en la regulación japonesa de contratos mediante IA de 2018, según el estudio comparativo complementario al estudio de impacto del Reglamento de IA: «Si bien las Directrices [*Guidelines*] reconocen la falta de una definición generalmente establecida de IA, ofrecen una clasificación "aproximada" en "(i) IA de uso general [*general-purpose AI*], basada en el concepto de crear máquinas que poseen inteligencia humana en sí misma ("IA fuerte"), y (ii) IA basada en el concepto de hacer que las máquinas realicen actividades que los humanos llevan a cabo con su inteligencia ("IA débil")». Cf. Japan

correspondencia directa con el lenguaje científico, sino con los objetivos para los que han sido creadas. En este caso, un sistema referido a un modelo que comprende de forma genérica distintos tipos de metodología —grandes modelos de lenguaje, modelos fundacionales, modelos de aprendizaje automático…—, unificados con relación a su función y características de uso, i.e. en relación con el modo de afectación de los derechos y obligaciones de los usuarios a partir de su introducción en el mercado.[41]

***Sin embargo, plantear la cadena de valor de ese modo significa asimilar todos los aspectos sociales, regulatorios y culturales a una única dimensión de mercado***. El dilema, pues, consiste en si (i) aplicar directamente el esquema clásico al diseño, refinamiento, puesta a punto, comercialización y mantenimiento de los sistemas inteligentes *como producto*, (ii) o bien redefinir esta aplicación para cada una de las dimensiones que son intangibles como sistemas de información con impacto efectivo en el entorno pero que *no son directamente monetizables,* puesto que no se basan en precio y dinero, i.e., en costes y beneficios económicos solamente.

La primera opción es la seguida por entidades financieras, servicios bancarios y reguladores económicos, siguiendo el esquema original de Porter con variaciones para adaptarlo a la complejidad tanto de la elaboración de los sistemas de IA como a sus efectos en la capacidad de innovación de las corporaciones y empresas. Así, según los analistas de la Dirección del Tesoro de Francia:

> La cadena de valor de la IA se halla segmentada en tres bloques principales: (i) los insumos necesarios para desarrollar sistemas y servicios de IA (capacidad de cómputo, datos, mano de obra especializada); (ii) el modelado, que incluye el desarrollo de modelos de IA de propósito general (modelos de base) y su especialización; y (iii) la implementación de estos modelos para los usuarios finales.[42]

Algunos analistas anticipan que la cadena de valor de la IA se basa en un ecosistema difícil de describir en su totalidad ya que abarca *hardware*, la nube, software, modelos de IA y aplicaciones verticales y de consumo. Un sistema muy complejo, con multitud de actores, que tardará años en desarrollarse, pero que cambiará el modelo de innovación y crecimiento del mercado.[43]

Voy a optar en estas páginas por la segunda opción, centrándome en la posibilidad de proyectar una cadena de valor ética y jurídica que hay que definir. Es decir, postulando requisitos como derechos, obligaciones y valores que no se definen solamente como componentes (actividades de soporte) de una cadena genérica sino como constituyentes ellos mismos de CVIA específicas que pueden alinearse con las CVIA económicas para crear valor añadido, si es el caso. No está

---

Ministry of Economy, Trade and Industry (METI), *Contract Guidelines on Utilisation of AI and Data* (2028), sintetizadas en Renda *et al.* (2021, pp. 85-86).

[41] Cf. *Ibid*. Art.3.63: un modelo de IA de uso general es «un modelo de IA, también uno entrenado con un gran volumen de datos utilizando autosupervisión a gran escala, que presenta un grado considerable de generalidad y es capaz de realizar de manera competente una gran variedad de tareas distintas, independientemente de la manera en que el modelo se introduzca en el mercado, y que puede integrarse en diversos sistemas o aplicaciones posteriores, excepto los modelos de IA que se utilizan para actividades de investigación, desarrollo o creación de prototipos antes de su introducción en el mercado».

[42] Chardon-Boucaud *et al.* (2024, p. 1). Los analistas indican el débil posicionamiento de la industria francesa respecto a la cadena de valor de la IA, la soberanía digital y la capacidad de competición. Señalan asimismo sus aspectos positivos : «*la France est mal positionnée sur le segment du calcul, mais dispose d'une main d'oeuvre qualifiée*». Cf. también, sobre el mismo tema, Besson *et al*. (2024) y Guthmann *et al*. (2024).

[43] Cf. Cuofaro (2025) precisa: «Se trata de un ecosistema en pleno desarrollo, donde más que atascos en la demanda, existen atascos en la oferta. Esto significa que el ecosistema aún está desarrollándose para cubrir una posible incorporación de IA en cada caso de uso comercial posible. Sin embargo, este proceso podría tardar un par de décadas en desarrollarse plenamente para dar cabida a la adopción masiva de la IA y poner la inferencia al alcance de todos».

excluido y es posible pensar en múltiples combinaciones en función del contexto y de los problemas a solucionar. Pero también podría considerarse, por ejemplo, un valor social o cultural con un coste que no produzca beneficios en el mercado sino en la sociedad civil, en los derechos humanos, o en la administración de los ciudadanos. Esto es lo que sucede habitualmente en servicios públicos como la sanidad, la educación, el transporte o la atención a minorías y grupos sociales vulnerables.

Es de notar que el mismo Porter se dio cuenta de la ventaja competitiva que implicaba la asunción de un marco de regulación adecuado para superar la tensión entre ecología y crecimiento económico. Es relevante la cita en este punto:

> *La posibilidad de que la regulación actúe* como estímulo para la innovación surge porque el mundo no se ajusta a la creencia panglosiana de que las empresas siempre toman decisiones óptimas. Esto solo será cierto en un marco de optimización estática donde la información es perfecta y ya se han descubierto oportunidades rentables para la innovación, de modo que las empresas con ánimo de lucro solo necesitan elegir su enfoque. Por supuesto, esto no describe la realidad. En cambio, el proceso real de competencia dinámica se caracteriza por oportunidades tecnológicas cambiantes, junto con información muy incompleta, inercia organizacional y problemas de control que reflejan la dificultad de alinear los incentivos individuales, grupales y corporativos. Las empresas tienen numerosas vías para la mejora tecnológica y una atención limitada. [44]

## 4. La cadena de valor de suministros (CVS), la cadena global de valor (CGV), la cadena de valor de la inteligencia artificial (CVIA)

Hay que distinguir algunos conceptos que se relacionan en el desarrollo de la cadena genérica de valor y su aplicación estratégica en diversos mercados (incluyendo el mercado global):

**1. Cadena de valor de suministros (*Value supply chain*):**

- *Definición*: Creación de valor en la red operativa para el abastecimiento y el suministro de productos a un cliente final.
- *Objeto*: Flujo físico de materiales y productos.
- *Actividades clave*: Abastecimiento, fabricación, distribución, atención al cliente.
- *Objetivos*: Optimización de la logística, gestión de inventarios, reducción de costes.

**2. Cadena global de valor (*Global value chain*):**

- *Definición*: Serie de actividades para comercializar un producto a escala global.
- *Objeto*: Todas las actividades para generar un producto, desde el diseño hasta la distribución global.
- *Actividades clave*: Preproducción (diseño), producción (fabricación), postproducción (marketing, distribución).
- *Objetivos*: Importación y exportación de bienes y servicios, gestión de redes de producción globales, garantía de beneficios equitativos para todos los países involucrados.

**3. Cadena de valor de la IA (*AI value chain*):**

- *Definición*: Marco estructural para convertir los datos en inteligencia y valor añadido en las dimensiones de mercado, social, cultural, ética y jurídica.

---

[44] Cf. Porter y van Linden (1995, p. 99). Esta es la perspectiva adoptada por los Informes Draghi y (especialmente) Letta, y por los analistas que incluyen la regulación o los mecanismos de gobernanza multinivel entre los componentes de la cadena de valor de la IA. Cf. Arnal (2025), Nannini (2025). Cf. también los Informes de la Comisión sobre la aplicación extensa de la IA en los ámbitos de la ciencia, la innovación y la industria, Heitor, Ferguson *et al.* (2024), Bianchini *et al.* (2025).

- *Objeto*: El ciclo de vida completo, desde la adquisición de datos hasta el acceso del usuario.
- *Actividades clave*: Adquisición de datos, entrenamiento de modelos, infraestructura, implementación, comercialización, monitorización.
- *Objetivos*: Mitigación de riesgos (sesgos, seguridad), garantía del cumplimiento de regulaciones fragmentadas, gestión de la responsabilidad distribuida, protección de la propiedad intelectual, delegación de autonomía, cadena de responsabilidad.

La inteligencia artificial ha añadido complejidad al concepto de cadena genérica de valor, puesto que puede integrarse en la «cadena de suministros» (CS)[45] y en la denominada «cadena global de valor» (CGV) o, aún más recientemente, en la «cadena de suministros de la infraestructura de la IA» (ISC). La *cadena de valor de la inteligencia artificial* constituye un marco que ilustra cómo las organizaciones crean valor al transformar datos brutos en inteligencia y, en última instancia, si cabe, en una ventaja competitiva. En mi opinión, el concepto posee asimismo una dimensión social, ética y jurídica que hay que representar. No se trata simplemente de una secuencia lineal de pasos, sino de la *creación de ecosistemas dinámicos e interdependientes*.

El marco ejerce una triple función: (i) como modelo industrial para empresas que desarrollan y ofrecen soluciones de IA; (ii) como dinamizador o facilitador (*enabler*) horizontal que impulsa la transformación en otros sectores como manufacturas, la logística, la atención médica, la movilidad o el turismo, abarcando los catorce espacios comunes de datos europeo; (iii) y como dinamizador de la producción de efectos sociales de mejora en los ecosistemas éticos, jurídicos y culturales que integran. La comprensión de este marco es esencial para identificar inversiones estratégicas, gestionar riesgos, garantizar el cumplimiento normativo y pensar al mismo tiempo la construcción de las sociedades híbridas contemporáneas (entre humanos y máquinas).[46]

Tres observaciones más. Por un lado, las cadenas de valor en IA no son equivalentes a las cadenas de suministros.[47] Éstas se organizan según una *lógica de bienes dominante*, lo que hace que sus resultados sean consumibles, mientras que las cadenas de valor siguen una *lógica de servicios* basada en la intangibilidad, los procesos de intercambio y una red más amplia de relaciones de colaboración y cocreación.[48] Sostengo que podemos extender esta idea de conectar las teorías de la ética y la gestión organizativa a nuestra idea de ecosistemas éticos y jurídicos integrados a través de la gobernanza jurídica de los modelos y sistemas inteligentes; es decir, abarcando la autonomía, la agencia moral, la agencia delegada, el daño y la responsabilidad.[49]

---

[45] La «cadena de suministros» (CS) (*supply chain*) es un concepto habitual en economía de la empresa. Se refiere genéricamente a la red de coordinación de todos los agentes, instalaciones y actividades comerciales necesarias para llevar un producto desde la materia prima hasta el cliente final.

[46] Los modelos de la cadena de valor de la IA a menudo son derivados a partir de la cadena de valor más específica del aprendizaje automático (ML), que consta de cinco etapas principales: *recopilación de datos, almacenamiento de datos, preparación de datos, entrenamiento de algoritmos y desarrollo de aplicaciones*. La IA generativa ha ampliado este modelo, añadiendo componentes posteriores más granulares. Por ejemplo, la parte final de la cadena de valor ahora incluye operaciones de aprendizaje automático (MLOps), que agilizan el proceso de llevar los modelos a producción, así como etapas diferenciadas para aplicaciones y servicios. Esta progresión desde conceptos fundamentales hasta aplicaciones más complejas del mundo real subraya la rápida evolución de la industria misma y, por ende, del impacto de la IA en el mercado digital.

[47] La cadena de suministros de LLMs ha sido recientemente definida de la siguiente manera: «La cadena de suministros se divide en tres capas principales: la capa de infraestructura del modelo (que incluye recursos computacionales, conjuntos de datos y cadenas de herramientas para entrenamiento, optimización e implementación), la capa de base del modelo (que abarca el entrenamiento, las pruebas, el lanzamiento y el mantenimiento continuo) y el ecosistema de aplicaciones posteriores (que facilita la integración de modelos preentrenados en diversas aplicaciones inteligentes)» (Wang *et al.* 2025, 147: 3).

[48] Engler y Renda (2022) definen la CVIA como «el proceso organizativo mediante el cual se desarrolla un sistema de IA individual y se pone en uso (o se despliega)». Attard-Frost y Gray Widder (2025) siguen esta línea organizativa y proponen una perspectiva integradora de la «ética que prioriza las cadenas de valor implicadas en el suministro de recursos y la recepción de resultados de los sistemas de IA, integrando una amplia gama de consideraciones éticas entre diversos actores, recursos, contextos y escalas de actividad». IOT Analytics relaciona directamente la CVIA con la inteligencia artificial generativa (GenAI), « una

Por otra parte, en segundo lugar, hay una relación entre la «cadena global de valor» (CGV) y la CVIA. Desde una perspectiva regulatoria, existen diferencias que justifican un tratamiento separado, pero también hay relaciones que deben considerarse. El concepto de CGV fue respaldado por la Organización de las Naciones Unidas para el Desarrollo Industrial (UNIDO) hace aproximadamente diez años, centrándose en la progresión económica y la comparación de los países en desarrollo.[50]

Finalmente, en tercer lugar, este proceso no se produce en un vacío regulatorio. La dimensión jurídica de las CGV implica que éstas no pueden tratarse de forma homogénea, ya que (i) se tornan jurisdiccionales, es decir, están sujetas al derecho positivo, tanto nacional como internacional, (ii) pasan a depender de la dinámica interna de las diferentes áreas jurídicas (derecho societario, derecho mercantil, derecho administrativo, etc.). Volvemos, pues, a la imagen del «mosaico» o de la «falda escocesa», ahora en el seno de las regulaciones de la UE.[51] Estas dificultades de coordinación constituyen algunas de las razones por las que prefiero distinguir la dimensión ética y jurídica de la CVIA para adecuarla luego a la económica antes que integrarla directamente o asimilarla a ésta. Si es así, el valor debería volver a definirse de nuevo, puesto que no estamos hablando solamente de beneficio económico y de competitividad.

---

técnica de aprendizaje profundo basada en autocodificadores variacionales, redes generativas adversarias y modelos basados en transformadores». Así, la «pila» de aplicaciones de la IA generativa incluye: La pila tecnológica de GenAI incluye cinco componentes: (i) Aplicaciones (p. ej., soluciones de software basadas en IA); (ii) herramientas de plataforma para implementación y gestión; (iii) modelos de base como GPT 4 de OpenAI; (iv) infraestructura crítica de *backend*, como procesamiento de datos y GPU; (v) marcos de gobernanza para la seguridad y el cumplimiento normativo. Cf. IOT Analytics (2025).

[49] El lector puede encontrar una exposición un poco más extensa de lo que sigue en Casanovas y Noriega (2025).

[50] UNIDO (2015) define las cadenas de valor según cuatro criterios o «puntos de entrada»: (i) como conjuntos de actividades que agregan valor; (ii) como matrices de vínculos; (iii) como redes o sistemas; (iv) como ciclos. Distingue entre enfoques positivos y normativos que deben combinarse. El primero cubre medios heurísticos convenientes para ayudar a comprender cómo funcionan, por ejemplo, cómo se fragmentan y distribuyen los pasos de producción entre ubicaciones geográficas y dimensiones analíticas como: (i) estructura de insumo-producto; (ii) distribución geográfica; (iii) el papel de las empresas líderes o intermediarios influyentes, proveedores, comerciantes, etc.; y (iv) el contexto institucional a nivel internacional, regional, nacional o local. Este último, normativo, se centra en lo que debe cambiar para mejorar el desempeño de la cadena de valor, las prioridades políticas vinculadas a los objetivos de desarrollo, los derechos, las obligaciones y los estándares. En nuestra opinión, lo importante en este enfoque es la concepción de la cadena de valor como proceso y resultado, es decir, como la combinación de diversas dimensiones en los ejes descriptivo y normativo que constituye adecuadamente un proceso de institucionalización.

[51] En el derecho de la UE, esto conlleva lo que Beckers (2023) ha denominado una institucionalización fragmentada de las cadenas globales de valor en al menos tres perspectivas: (i) actor individual (derecho societario); (ii) actor colectivo (derecho del consumidor); (iii) institucionalización despersonalizada (derecho mercantil y comercial).[51] Así, sugiere «la imagen jurídica como fragmentada en diferentes instituciones jurídicas (empresa, red, mercado) que se relacionan *con diferentes áreas jurídicas (derecho societario, derecho del consumidor, prácticas de mercado y derecho mercantil) y se correlacionan con diferentes objetivos políticos (sostenibilidad, digitalización, resiliencia, protección del consumidor, equidad)* [énfasis nuestro]». La naturaleza compleja y multipartita de la cadena de valor de la IA presenta un desafío fundamental para los marcos jurídicos tradicionales. El «problema de ritmo» abordado por Beckers (2023) —la incapacidad de la ley para mantenerse al día con la rápida evolución de la tecnología— es evidente en los intentos de aplicar los principios legales establecidos al daño inducido por la IA.

## 5. Obstáculos, supuestos y propuestas de superación

Podríamos hacer hincapié además en que las características de una única CVIA no son siempre positivas. Por ejemplo, la concentración de poder económico y político en las capas fundamentales de la cadena de valor de la IA implica que las tecnologías de IA suelen ser desarrolladas y controladas por grandes corporaciones, denominadas *gobernantes de la cadena de valor* (*value chain governors*) que actúan como «cuasiestados».[52] Esto crea una nueva dinámica donde la facilidad con la que estos gobernantes pueden reubicar su producción en países con menores costos representa un riesgo. El modelo tradicional de la cadena de valor se complica aún más por el ecosistema de IA más amplio que lo sustenta. Este ecosistema incluye componentes fundamentales de desarrollo, en particular *hardware informático y plataformas en la nube*.

La capa de hardware está dominada por unos pocos gigantes de los semiconductores, como Nvidia, Intel y AMD, manteniendo Nvidia una posición particularmente dominante en el mercado de *chips* de IA. De igual manera, las plataformas en la nube —quienes proporcionan la infraestructura informática escalable necesaria para el desarrollo de la IA— están controladas por unos cuantos *hiperescaladores* como Amazon Web Services (AWS), Microsoft Azure y Google Cloud. ¿Cómo enfrentarse a este reto? Podemos partir de algunos supuestos que presuponen la posibilidad de diseñar CVIA independientes:

1) ***El diseño de la CVIA en contextos híbridos entre agentes humanos y artificiales***

    Esto significa que podemos reestructurarlo en nuevos marcos, donde la red de datos, el internet de las cosas y la llamada Industria 4.0 (y 5.0) añaden complejidad al mismo, puesto que abren nuevas dimensiones en el uso de los conceptos jurídicos (como el de validez ecológica de las regulaciones). En un artículo reciente, hemos desarrollado un marco cuatridimensional para la validez de los ecosistemas jurídicos —*lenguaje, sociedad, legalidad y datos*— utilizando un modelo de hipercubo (teseracto).[53] *La validez de los ecosistemas jurídicos puede alinearse con las cadenas de valor de la AI que diseñan los modelos para implementar los sistemas*. Los mecanismos cambian, pero la idea básica de mantener un equilibrio proporcional entre los intereses en la construcción de una sociedad justa y equitativa se mantiene. La figura 2 representa un ejercicio de relación a partir de una noción de validez jurídica compatible con la CVIA (en el centro). No se trata de prescindir de los constructos teóricos del derecho ya existentes (ley, norma, eficiencia, validez, efectividad, etc.), sino de completarlos mediante conceptos intermedios adaptados a las dimensiones y entornos dinámicos de la web de datos e internet de las cosas.

---

[52] Li (25) ha mostrado como las grandes corporaciones tecnológicas imponen su gobernanza mediante: (1) la incorporación de parámetros de viabilidad técnica en regulaciones y estándares formales mediante el despliegue estratégico de experiencia; (2) el establecimiento de una autoridad normativa desplegada mediante innovaciones autorreguladoras que se difunden globalmente a través de plataformas multilaterales; y (3) la configuración de marcos de significado compartidos que definen cómo se conceptualizan y abordan los retos de gobernanza. Esta noción tiene su origen en la de *global governors* (gobernantes globales), definidos como «autoridades que ejercen poder transfronterizo para influir en las políticas públicas. Por lo tanto, los gobernantes plantean problemas, definen agendas, establecen e implementan normas o programas, y evalúan y/o adjudican resultados». (Avant *et al.* 2010, p. 2)

[53] Casanovas (2025b).

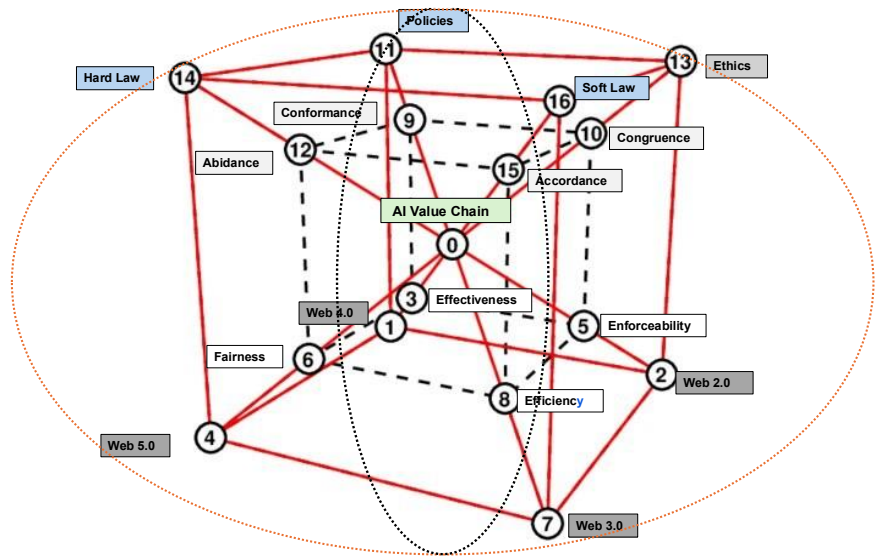

**Fig. 2.** Hipercubo para la representación de conceptos jurídicos en cuatro dimensiones (lenguaje, sociedad, legalidad y datos). Las elipsis representan las aproximaciones epistemológicas *middle-out* (del medio afuera) e *inside-out* (de dentro afuera) internas a las plataformas, aplicaciones y sistemas, en lugar de las representaciones verticales y horizontales externas del derecho. Fuente: Casanovas (2025b).

2) ***La redefinición del concepto de validez jurídica desde un punto de vista empírico, relacionado con la CVIA***

Hay que insistir en dos requerimientos compatibles: (i) la necesidad de mantener la estructura ya existente a partir de los conceptos jurídicos básicos, ya mencionados en el primer punto; (ii) la necesidad de completarlos adoptando una perspectiva basada en los instrumentos y procesos de *implementación* de derecho —gobernanza jurídica— desde la base antes que una *aplicación* del derecho exclusivamente vertical. La necesidad de mantener los conceptos jurídicos básicos, especialmente respecto a la responsabilidad extracontractual (*torts*), ya ha sido señalada en tecnología.[54] Algunos de estos conceptos son antiguos y provienen de la tradición de derecho romano.[55] Lo que nos gustaría apuntar aquí es que resulta posible equiparar la posición de la cadena de valor de la IA con la posición que ocupa la *validez normativa*—i.e. la noción de *legalidad*— en el sistema jurídico respecto a otras propiedades como la eficacia y la efectividad. Validez y CVIA son conceptos de segundo orden que poseen una dimensión regulatoria formal y otra empírica. Por la primera, pueden ser aplicadas, por ejemplo, a cadenas de responsabilidad que deben ser ajustadas a derecho; por la segunda, pueden ser aplicadas a reglas que responden a distinto grado de cumplimiento mediante el establecimiento de indicadores y *umbrales* (*thresholds*). Esto implica una aproximación empírica. Dicho de otro modo: es posible establecer métricas que indiquen la relación entre los conceptos y el grado de cumplimiento normativo (*compliance*). Así, la cadena de valor de la IA puede medirse no solamente en términos económicos, sino como el valor que la IA aporta en cada nivel de gobernanza para la emergencia y mantenimiento sostenible de ecosistemas regulatorios que resultan aceptados y puestos regularmente en práctica por los participantes interesados o actores del sistema *(stakeholders)* (vid. *infra*, punto 5).

---

[54] Leenes (2019), Soyer y Tettenborn (2022).
[55] Cf. Pagallo (2013, p. 82) sugiere que la idea que algunos tipus de robot pueden ser considerados como responsables de su comportamiento tiene precedentes en la figura del *peculium* contemplada en el Digesto, mediante la cual los esclavos, que carecían de personalidad jurídica y de las capacidades que la acompañaban, podían actuar como gestores de la propiedad, banqueros o comerciantes. Cf. asimismo Rodríguez de las Heras (2023) sobre la figura de los contratos *intuitu hominis* (en relación a la específica personalidad del contratante como fundamento del contrato).

3) *La elusión del incremento de complejidad en el análisis de riesgos*

No hay que incrementar la complejidad en el análisis de riesgos, sino más bien reducirla. No tiene sentido añadir más complejidad a un conjunto de relaciones que ya de por si resultan difíciles de definir y analizar. En cambio, sí que debemos afrontar al menos cuatro cuestiones técnicas que aún no se han resuelto satisfactoriamente: (i) el problema de imbuir valores en los sistemas de inteligencia artificial (problema de alineación de valores, VAP);[56] (ii) el problema de extraer reglas formales de las normas expresadas en lenguaje natural;[57] (iii) el problema de integrar las tecnologías de IA simbólicas y generativas desde una aproximación neurosimbólica para la adquisición de conocimiento y su representación;[58] (iv) los problemas existentes en lógica deóntica —como la del permiso débil— para formalizar el razonamiento ético y jurídico.[59]

4) *La elaboración de esquemas simples de los que partir para el análisis de la gobernanza de los sistemas de inteligencia artificial*

La Figura 3 presenta un esquema mínimo inicial que se centra únicamente en las dos ramas principales de la autonomía de los sistemas inteligentes: la *autonomía delegada* y la *agencia moral*. Es muy probable que este esquema se modifique en un futuro próximo, pero, no obstante, resulta útil mostrar que las formulaciones simples pueden ayudar a resolver problemas complejos. Cabe destacar que, en el centro del diagrama, hemos situado el problema principal: la regulación y el control de los sistemas de inteligencia artificial a través de la propia inteligencia artificial. La intervención humana, especialmente en sistemas de IA generativos, no puede obviarse. Debe combinarse con técnicas de IA. La intervención humana en el ciclo de vida del sistema (*human-in-the-loop*) implica también la intervención de la IA en el circuito.[60]

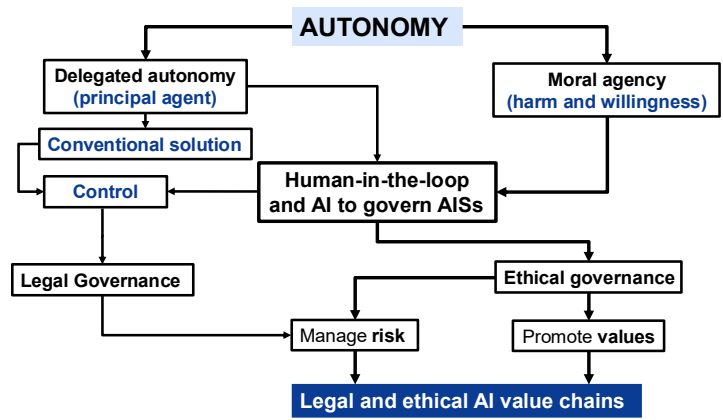

**Fig. 3.** Esquema inicial para la gobernanza de los sistemas de inteligencia artificial. Fuente: Casanovas y Noriega (2025).

---

[56] Cf. Noriega *et al*. (2023), Osman (2024).
[57] Este problema, descrito en Hashmi *et al*. (2018), aún no ha sido completamente solucionado.
[58] Cf. entre muchos otros, Hitzler *et al*. (2022). El enfoque neurosimbólico, que hibrida los aspectos simbólicos y subsimbólicos, «se basa en la inyección de conocimiento procedente de grafos de conocimientos sobre los modelos de lenguaje ya entrenados para reforzarlos y realizar un ajuste fino [*fine tuning*]» (Gómez Pérez 2023, p. 71).
[59] Cf. Governatori y Rotolo (2025).
[60] Cf. Casanovas y Noriega (2025). Véase también, sobre el tratamiento de riesgos en la gobernanza de sistemas inteligentes, Noriega y Casanovas (2025).

Ha habido hasta ahora una ingente cantidad de contribuciones respecto a la agencia moral de los sistemas de IA (especialmente de los sistemas multiagente). No vamos a entrar en ella, puesto que la mayoría de las discusiones se han producido a nivel teórico (o filosófico) sobre los conceptos de autonomía y moralidad.[61] Como veremos en el apartado siguiente, la perspectiva epistemológica adoptada no requiere posicionarse sobre los sistemas como sujetos morales, sino sobre la autonomía y los conceptos éticos y jurídicos que permiten su gobernanza, incluyendo la minimización de riesgos y la respuesta en caso de producción de daños o efectos inesperados o no deseados.[62] Se sitúa en la consideración de los sistemas multiagente como *artefactos* basados en instituciones electrónicas o en línea (*electronic / on-line institutions)*, y de los sistemas de IA generativa (LLM, FM, DL) como *instrumentos* basados en la teoría matemática de procesos estocásticos (con el precedente de las redes neuronales) para el tratamiento del lenguaje.

### 5) *El diseño de ecosistemas éticos y jurídicos*

Como resultado, el enfoque empírico escogido permite incentivar y promover la aparición de ecosistemas regulatorios que incorporan las cadenas de valor ética y jurídica de la IA (CVIA). Su objetivo es la materialización de valores en la organización, coordinación e implementación de los sistemas autónomos inteligentes (o sistemas multiagente) y de la IA generativa *en contextos y plataformas específicos*. Este campo está experimentando un rápido desarrollo, especialmente desde que el concepto se ha incorporado en varios considerandos y artículos de la legislación europea reciente, constituyendo una guía para la mitigación de riesgos, la promoción de los valores éticos y el desarrollo de protecciones basadas en los derechos humanos. Pero dejando aparte el desarrollo legislativo, existen desde el punto de vista de la computación algunas propuestas basadas en una aproximación en abstracto a la dimensión ética en las interacciones y los flujos de información. Subrayan consistentemente la importancia de una perspectiva de *escala* y la distinción entre los distintos *niveles* y *grados* en la interacción entre humanos, robots y sistemas de información (Human-Machine Interaction, HMI).[63] Las interacciones, a su vez, pueden ser formalizadas.[64] Lo que debemos subrayar es que la formalización también es (por lo menos parcialmente) aplicable en la coordinación de los distintos módulos y niveles de procesamiento de información en las plataformas, tal como hemos efectuado ya en el campo de las manufacturas inteligentes (*smart manufacturing*). La CVAI puede ser, pues, más compleja y extensa de lo que hasta ahora ha sido considerada.[65]

---

[61] Fritz *et al.* (2020) clasifican las teorías éticas sobre los agentes como sujetos morales en tres categorías: (i) *tecnocéntricas* (*technological agency*, e.g. Luciano Floridi); (ii) *antropocéntricas* ( e.g., *triadic agents* Deborah Johnson); (iii) *constructivistas* (*hybrid agency*, e.g. Peter-Paul Verbeek).

[62] Casanovas y Noriega (2025) contemplan los conceptos de autonomía (*legal autonomy*), gobernanza (*legal governance*), daño (*harm*), y responsabilidad jurídica (*liability*). El capítulo contiene una crítica de los sistemas de IA como sujetos jurídicos y de la denominada teoría de la red de agentes (*Actor-network theory*, ANT).

[63] Gabriel *et al.* (2020, 190) subrayan que una evaluación comprensiva de la ética de los asistentes de IA requiere un análisis de las siguientes tres capas: (i) las capacidades de la IA, midiendo resultados de salida y las funciones del sistema o de sus componentes (los datos para su entreno, e.g.); (ii) la interacción H/M, con la medición de los riesgos de dañar a una persona que interaccione con el sistema; (iii) sistemas más amplios, con la medición de riesgos de daños mediante análisis ambientales o económicos. Para una aproximación basada en el tipo de interacción, cf. V. Dignum (2019). Para una consideración de la responsabilidad en los sistemas multiagentes (SMA) como sistemas sociotécnicos, *vid*. Yazdanpanah *et al.* (2021): «Para que los agentes de un SMA puedan razonar sobre la responsabilidad, es necesario distinguir diferentes formas de responsabilidad y articular cómo cada una se relaciona conceptualmente con la capacidad estratégica (y la distribución del poder); la capacidad epistémica (y la distribución del conocimiento); las tareas (y la distribución de las obligaciones); y las normas y valores (y la distribución de las preferencias)».

[64] Cf. F. Dignum (2018).

[65] Para construir ecosistemas jurídicos, i.e. sistemas regulatorios incorporados (*imbued*) en las arquitectura y procesos de información (*workflows*) de las plataformas, hay que considerar sus diversas dimensiones y etapas, desde el sistema en la nube hasta el trabajador, y de ese micronivel de producción al de distribución,

*6) El diseño de la dimensión ética y jurídica de la cadena de valor*

Una vez establecidos los contextos y escenarios posibles dentro y fuera de las plataformas, habría que plantear la compatibilidad entre la selección de valores que se implementan, los requerimientos procedentes de la interpretación de las diferentes disposiciones normativas, y la cadena de valor de la IA en el mercado. No hay un único modo de realizar esta operación, y tampoco puede obviarse la función económica de la cadena de valor para las empresas y corporaciones tecnológicas. En este sentido, los investigadores del Ada Lovelace Institute relacionaron tempranamente, cuando aún se estaba discutiendo la versión final del Reglamento europeo de IA, la cadena de valor de la IA con las estrategias de mercado de la economía de plataforma: «hay dos formas principales en que los sistemas GPAI [*general-purpose AI*] y sus modelos subyacentes se hacen accesibles a los desarrolladores posteriores en el mercado: a través de API [*application programming interface*] y mediante acceso de código abierto».[66] Corresponden a dos modos distintos de monetización y de relación con los usuarios. En la primera, el control sobre el modelo y el Código permanece en manos del proveedor. En la segunda, el modelo o algunos de sus elementos se sitúan en el dominio público y se permite su modificación y distribución según los términos que establezca la licencia.

*7) La cadena de valor ética y jurídica como incentivo de gobernanza financiera*

Finalmente, puede subrayarse el carácter incentivo que la CVIA posee en los espacios comunes de datos para los instrumentos financieros. En contraste con la nueva política monetarista norteamericana que apuesta al mismo tiempo por las criptomonedas y criptoactivos sin apenas restricción y la defensa de la posición central del dólar —*i.e.*, por una versión directamente económica, privada y corporativa de la CVIA[67]—, la UE apuesta por la defensa de los derechos de los inversores, consumidores y ciudadanos en el espacio de datos comunes financieros. La regulación europea pretende conseguir, a grandes rasgos, que los datos financieros sean compartidos mediante la creación de «finanzas abiertas» (*Open finance*)[68] y el establecimiento de medidas de protección incluidas en el *Reglamento de mercados de criptoactivos* (MiCA).[69] Según la OECD (2023) la práctica de compartir datos de empresas financieras permitiría circunscribir mejor la situación financiera de los usuarios y obtener una mayor interoperabilidad entre sus distintos ámbitos. Por su parte, el Reglamento define los criptoactivos como «una representación digital de un valor o de un derecho que puede transferirse y almacenarse

---

hasta que llega al consumidor. Cf. Margetis *et al.* (2021), Margetis et al. (2022), Casanovas y Hashmi (2024), Casanovas (2024), Casanovas *et al.* (2025).

[66] Cf. Küspert, Moes, y Dunlop (2023).

[67] Es decir, la política financiera de apoyo de los activos digitales y *blockchain* en todos los sectores de la economía con los siguientes objetivos: «(i) Proteger y promover la capacidad de los ciudadanos y las entidades del sector privado para acceder y utilizar con fines lícitos las redes públicas de cadena de bloques (*blockchain*) abiertas sin persecución, incluyendo la capacidad de desarrollar e implementar *software*, participar en la minería y validación, realizar transacciones con otras personas sin censura ilegal y mantener la autocustodia de los activos digitales; (ii) promover y proteger la soberanía del dólar estadounidense, incluyendo acciones para promover el desarrollo y el crecimiento de las monedas estables legales y legítimas respaldadas por el dólar en todo el mundo». Cf. US Presidential Action (2025, 23 enero) sobre tecnología digital financiera.

[68] «Open Finance» pretende introducir «un sistema de compensación para los titulares de datos, imponiendo requisitos de estandarización, estableciendo esquemas de intercambio de datos financieros para desarrollar mecanismos de coordinación dentro de la industria e introduciendo paneles de permisos, para que los clientes monitoreen sus permisos de datos» (Pastor Sempere, 2023, p. 26).

[69] MiCA (mayo, 2023). Por ejemplo, la protección de titulares y clientes de proveedores de servicios de criptoactivos y la creación de medidas para prevenir el uso de información privilegiada, la divulgación ilícita de información privilegiada y la manipulación del mercado. Para un análisis de los mecanismos y sus limitaciones, de los Artículos 86 al 92, *vid*. Echebarría Sáenz (2025).

electrónicamente, mediante la tecnología de registro distribuido [*distributed ledger technology*] o una tecnología similar».[70] Distingue entre criptodivisas o criptomonedas y *tokens*. En todo caso, consiste en la representación digital de un valor, i.e., en datos que pueden almacenarse, operarse, transferirse y convertirse asimismo en inteligencia. La CVIA, en este caso, proviene de la seguridad, transparencia y trazabilidad lograda en los diversos niveles, independientemente de su dimensión pública o privada. Se trata de una gobernanza descentralizada que facilita (como *enabler*) la transmisión y creación de valor en las transacciones. De ahí la importancia de los espacios comunes de datos para contextualizarlas y regenerar el espacio público,[71] así como las ideas procedentes del *Semantic Web Consortium* (W3C) para redefinir la identidad digital en el denominado «paradigma de la identidad digital auto-soberana» (*Self-Sovereign Identity paradigm*), i.e. dirigida a a la gestión de datos sin depender de una autoridad central.[72]

## 6. El Reglamento Europeo de Inteligencia Artificial

Vayamos ahora al Reglamento, a la denominada Ley Europea de la IA.[73] En los documentos previos de carácter normativo se subrayaban los valores éticos que deben preservarse en la cadena de valor y la cadena de responsabilidades que debía dibujarse. Por ejemplo, el documento titulado *Marco de los aspectos éticos de la inteligencia artificial, la robótica y las tecnologías conexas* (2020) estipula los principios de necesidad y proporcionalidad[74], así como la protección de ciudadanos y consumidores. El Parlamento:

> 32. Sostiene que los valores éticos de la equidad, la exactitud, la confidencialidad y la transparencia deben ser la base de estas tecnologías, lo que en este contexto implica que sus operaciones deben concebirse de tal manera que no generen resultados sesgados;
>
> 78.Subraya la importancia de que haya un marco regulador para la inteligencia artificial que se aplique en aquellos casos en que los consumidores, dentro de la Unión, sean usuarios de un sistema algorítmico, estén sujetos a tal sistema, sean los destinatarios de tal sistema o estén orientados a él, independientemente del lugar en que estén establecidas las entidades que desarrollen, comercialicen o utilicen dicho sistema; estima además que, en aras de la seguridad jurídica, las normas establecidas en dicho marco deben aplicarse a todos los desarrolladores y a toda la cadena de valor, a saber, el desarrollo, el despliegue y el uso de las tecnologías pertinentes y sus componentes, y deben garantizar un alto nivel de protección de los consumidores.[75]

---

[70] MiCA (mayo, 2023). Art. 3.5.
[71] Cf. Casanovas (2025a).
[72] Cf. Rodríguez-Doncel (2025, p. 122): «La mayor parte de la población mundial posee al menos una identidad digital. Sin embargo, el concepto de identidad digital va mucho más allá de la autenticación de seres humanos en servicios en línea. La identidad es algo más importante que una simple invitación. La identidad es un sentido de identidad; se trata de cómo nos percibimos a nosotros mismos, nuestros valores, creencias, experiencias y relaciones; se trata de la comprensión interna de quién somos. Ahora vivimos en el mundo digital. Nuestros recuerdos ya no son incorpóreos y, una vez transformados en datos, pueden ser procesados y utilizados por algoritmos».
[73] EU Commission (June, 2024). Para un análisis de sus distintas fases de elaboración, con especial atención a la responsabilidad, véase Rodríguez de las Heras (2022) (2025a).
[74] Así, el Parlamento: «4.Considera que toda medida legislativa relativa a la inteligencia artificial, la robótica y las tecnologías conexas debe respetar los principios de necesidad y de proporcionalidad; 5.Estima que un enfoque de este tipo permitirá a las empresas introducir productos innovadores en el mercado y crear nuevas oportunidades, al tiempo que se garantiza la protección de los valores de la Unión, conduciendo al desarrollo de sistemas de inteligencia artificial que incorporen los principios éticos de la Unión desde el diseño; considera que un marco regulador de este tipo basado en valores representaría un valor añadido al aportar a la Unión una ventaja competitiva única y contribuiría de forma significativa al bienestar y la prosperidad de los ciudadanos y las empresas de la Unión impulsando el mercado interior; subraya que tal marco regulador para la inteligencia artificial representará también un valor añadido en lo que respecta a la promoción de la innovación en el mercado interior; opina que, por ejemplo, en el sector del transporte, este enfoque brinda a las empresas de la Unión la oportunidad de convertirse en líderes mundiales en este ámbito.» EU Commission (2020, p. 68).

En los documentos sobre responsabilidad civil, también se subrayaba la incidencia de la corresponsabilidad sobre la cadena de valor y la cadena de suministros.[76] La «responsabilidad compartida» debía estar en consonancia con los valores y distribuirse a lo largo de toda la cadena. Así se habían pronunciado los expertos en derecho civil desde hacía tiempo, estableciendo que «los robots impulsados por IA en espacios públicos deberían estar sujetos a una responsabilidad objetiva [*strict liability*] por los daños resultantes de su funcionamiento», señalando, por cierto, que «debemos resistirnos a las peticiones de establecer una personalidad jurídica basadas en la ciencia ficción».[77]

En la Propuesta de Ley, el concepto de cadena de valor había sido también asumido como central, casi como espina dorsal (i) de la integración en el mercado de los sistemas de inteligencia artificial, (ii) de la protección de los derechos de los ciudadanos, (iii) de la cadena de obligaciones y requisitos que deben cumplir los sistemas de IA que se comercialicen. En un principio, la Exposición de Motivos de la Propuesta de Ley vinculaba directamente la cadena de valor con los derechos fundamentales:

> Sirviéndose de un conjunto de requisitos destinados a conseguir que la IA sea fiable y que se impongan obligaciones proporcionadas a todos los participantes en la cadena de valor, la propuesta reforzará y promoverá la protección de los derechos salvaguardados por la Carta: el derecho a la dignidad humana (artículo 1), el respeto de la vida privada y familiar y la protección de datos de carácter personal (artículos 7 y 8), la no discriminación (artículo 21) y la igualdad entre hombres y mujeres (artículo 23). Su objetivo es evitar un efecto paralizante sobre los derechos a la libertad de expresión (artículo 11) y de reunión (artículo 12), y garantizar el derecho a la tutela judicial efectiva y a un juez imparcial, la presunción de inocencia y los derechos de la defensa (artículos 47 y 48), así como el principio general de buena administración.[78]

Inmediatamente, en la versión final del Reglamento, la expresión se convierte en *cadena de valor de la IA*. Así, en el título 1, «se proporciona una *definición clara de los principales participantes*

---

[75] EU Commission (2020, octubre, p. 68).
[76] «En el marco de la Unión en materia de seguridad de los productos, independientemente de la complejidad de la cadena de valor, la responsabilidad de la seguridad del producto recae en el productor que lo comercializa. Los productores son responsables de la seguridad del producto final, incluidas las partes integradas en el mismo, por ejemplo, los programas informáticos de un ordenador. […] La normativa de la Unión en materia de seguridad de los productos *tiene en cuenta la complejidad de las cadenas de valor e impone obligaciones a una serie de agentes económicos en consonancia con el principio de «responsabilidad compartida»* [énfasis añadido]. Si bien la responsabilidad del productor respecto de la seguridad del producto final ha resultado ser adecuada para las cadenas de valor complejas actuales, contar con disposiciones explícitas que pidan específicamente la cooperación entre los agentes económicos de la cadena de suministro y los usuarios puede aportar seguridad jurídica quizás hasta en cadenas de valor más complejas. En particular, cada agente de la cadena de valor que influya en la seguridad del producto (por ejemplo, los productores de programas informáticos) y los usuarios (al modificar el producto) asumirían su responsabilidad y proporcionarían al siguiente agente de la cadena la información y las medidas necesarias.» EU Commission (2020, febrero, p. 13).
[77] EU Expert Group on Liability and New Technologies (2019, p. 3 y p. 13. La tesis n. 11 subraya: «Si hay dos o más operadores, en particular: (a) la persona que principalmente decide y se beneficia del uso de la tecnología pertinente (operador de interfaz, *frontend*) y (b) la persona que define continuamente las características de la tecnología pertinente y proporciona soporte esencial y continuada (operador de subestructura, *backend*), la responsabilidad objetiva debe recaer en quien tenga mayor control sobre los riesgos de la operación». *Ibid.* p. 6.
[78] Propuesta de Reglamento del Parlamento Europeo y del Consejo (2021), 3.5 Derechos Fundamentales, p. 12.

*en la cadena de valor de la IA*, como los proveedores y los usuarios de sistemas de IA, entre los que se incluyen los operadores públicos y privados para garantizar la igualdad de condiciones».

En el texto definitivo, la idea se amplía para cubrir todas las dimensiones y niveles de la generación de sistemas de IA, su introducción en el mercado, la pirámide de riesgos y conceptos relacionados —como el de cumplimiento normativo— con el proceso global. En el considerando 83[79] y el 88[80] son explícitos respecto a la centralidad de la cadena de valor tanto en el ciclo de producción e implantación de los sistemas de IA como en su calidad de componentes de otros sistemas.

Especialmente el Considerando 101 establece:

> Los proveedores de modelos de IA de uso general tienen una función y una responsabilidad particulares a lo largo de la cadena de valor de la IA, ya que los modelos que suministran pueden constituir la base de diversos sistemas de etapas posteriores, que a menudo son suministrados por proveedores posteriores que necesitan entender bien los modelos y sus capacidades, tanto para permitir la integración de dichos modelos en sus productos como para cumplir sus obligaciones en virtud del presente Reglamento o de otros reglamentos. Por consiguiente, deben establecerse medidas de transparencia proporcionadas, lo que incluye elaborar documentación y mantenerla actualizada y facilitar información sobre el modelo de IA de uso general para su uso por parte de los proveedores posteriores. El proveedor del modelo de IA de uso general debe elaborar y mantener actualizada la documentación técnica con el fin de ponerla a disposición, previa solicitud, de la Oficina de IA y de las autoridades nacionales competentes. Los elementos mínimos que debe contener dicha documentación deben establecerse en anexos específicos del presente Reglamento. La Comisión debe estar facultada para modificar dichos anexos mediante actos delegados en función de los avances tecnológicos.[81]

Así pues, la Ley de IA de la UE menciona la CVIA desde el principio, en varios considerandos y artículos (especialmente el artículo 25: *Responsabilidades a lo largo de la cadena de valor de la IA*[82]; y en el artículo 56: *Código de Buenas Prácticas*).[83] Esta perspectiva se ha proseguido en los tres capítulos del *Código de buenas prácticas para los proveedores de modelos de IA de uso general*[84] y, especialmente, en las *Directrices sobre el alcance de las obligaciones de los modelos de IA de propósito general*, que vincula directamente la posición de las CVAIs con la cadena de responsabilidad que conllevan.[85]

---

[79] Considerando 83: «Teniendo en cuenta la naturaleza y la complejidad de la cadena de valor de los sistemas de IA y de conformidad con el nuevo marco legislativo, es esencial garantizar la seguridad jurídica y facilitar el cumplimiento del presente Reglamento. Por ello es necesario aclarar la función y las obligaciones específicas de los operadores pertinentes de toda dicha cadena de valor, como los importadores y los distribuidores, que pueden contribuir al desarrollo de sistemas de IA. En determinadas situaciones, esos operadores pueden desempeñar más de una función al mismo tiempo y, por lo tanto, deben cumplir de forma acumulativa todas las obligaciones pertinentes asociadas a dichas funciones. Por ejemplo, un operador puede actuar como distribuidor e importador al mismo tiempo». Reglamento (UE) 2024/1689) del Parlamento Europeo y del Consejo (2024, p. 23).

[80] Considerando 88: «A lo largo de la cadena de valor de la IA, numerosas partes suministran a menudo no solo sistemas, herramientas y servicios de IA, sino también componentes o procesos que el proveedor incorpora al sistema de IA con diversos objetivos, como el entrenamiento de modelos, el reentrenamiento de modelos, la prueba y evaluación de modelos, la integración en el software u otros aspectos del desarrollo de modelos. Dichas partes desempeñan un papel importante en la cadena de valor en relación con el proveedor del sistema de IA de alto riesgo en el que se integran sus sistemas, herramientas, servicios, componentes o procesos de IA, y deben proporcionar a dicho proveedor, mediante acuerdo escrito, la información, las capacidades, el acceso técnico y demás asistencia que sean necesarios habida cuenta del estado actual de la técnica generalmente reconocido, a fin de que el proveedor pueda cumplir íntegramente las obligaciones establecidas en el presente Reglamento, sin comprometer sus propios derechos de propiedad intelectual e industrial o secretos comerciales». *Ibid.* p. 24.

[81] *Ibid.* p. 27.

[82] Según el art. 25.1, cualquier distribuidor, importador, responsable del despliegue o tercero será considerado proveedor de un sistema de IA de alto riesgo (se entiende que en la «cadena de valor de la IA», concepto que no queda específicamente definido).

[83] Según el art. 56.2.d: La Oficina y el Consejo de IA velarán para que los códigos de buenas prácticas incluyan «las medidas, procedimientos y modalidades de evaluación y gestión de los riesgos sistémicos a escala de la Unión, incluida su documentación, que serán proporcionales a los riesgos y tendrán en cuenta su gravedad y probabilidad y las dificultades específicas para hacerles frente, habida cuenta de la manera en que dichos riesgos pueden surgir y materializarse a lo largo de la cadena de valor de la IA».

La cuestión es cómo. ¿De qué forma se compatibiliza la protección de derechos, la concatenación de obligaciones, el cumplimiento de requisitos formales y procesales, y lo que parece más importante para el Reglamento, la construcción del mercado digital europeo? ¿Y con qué métricas va a operarse para medir y validar los resultados? ¿Cómo se alinea la cadena de valor económica con la jurídica y ética? ¿Y qué papel debe cumplir el desarrollo de intangibles como la lengua, la cultura? ¿Y cómo va a medirse?

Estos problemas no están resueltos.[86] Hay una mezcla entre las cadenas de valor para la IA que hemos identificado en los puntos anteriores. Los documentos más específicos respecto a la CVAI — i.e. la evaluación de impacto del Reglamento de IA[87] y el estudio adicional que le asiste[88] — poseen esta característica, puesto que al mismo tiempo que la definen a lo largo del ciclo de vida e implementación de los sistemas, la asocian con la producción de riesgos y su mitigación para proteger los derechos de los ciudadanos.[89]

---

[84] EU Commission (10 de Julio de 2025), en los capítulos de Transparencia (*Transparency*), y de Seguridad y Protección (*Safety and Security*). En este último, Considerando 2 (a): «*Principio de Gestión Adecuada del Ciclo de Vida*. Los Firmantes reconocen que los proveedores de modelos de IA de propósito general con riesgo sistémico deben evaluar y mitigar continuamente los riesgos sistémicos, adoptando las medidas adecuadas a lo largo de todo el ciclo de vida del modelo (incluso durante el desarrollo previo y posterior a su comercialización), cooperando y teniendo en cuenta a los actores relevantes de la cadena de valor de la IA […]».

[85] EU Commission (17 de julio 2025), «*Antecedentes y objetivos de las directrices de la Comisión*. (2) Los modelos de IA de propósito general desempeñan un papel fundamental en la innovación y la adopción de la IA en la Unión. Esto se debe a que pueden utilizarse para diversas tareas e integrarse en una amplia gama de sistemas de IA posteriores. Por lo tanto, los proveedores de modelos de IA de propósito general desempeñan un papel y una responsabilidad específicos a lo largo de la cadena de valor de la IA, incluso respecto a los proveedores posteriores, quienes necesitan un buen conocimiento de los modelos y sus capacidades».

[86] Para un análisis de los documentos previos y la redacción final del Reglamento, vid. Rodríguez de las Heras (2025). La autora también analiza los proyectos legislativos sobre responsabilidad extracontractual. «La versatilidad de los modelos de IA de uso general, incluso al realizar tareas no entrenadas o inesperadas, abre un amplio abanico de posibles escenarios de responsabilidad. La Ley de IA no proporciona un régimen jurídico completo para la IA generativa ni identifica todos los escenarios de riesgo. Por el contrario, a medida que la IA generativa se extiende a las actividades económicas y sociales, los riesgos se diversifican y los casos de responsabilidad se intensifican. Desde casos penales hasta situaciones de responsabilidad contractual, desde infracciones de la privacidad o los derechos de autor hasta diversas violaciones de derechos fundamentales, la generalidad, la multimodalidad y el carácter fundacional de los modelos de IA de propósito general despliegan una cadena de infracciones de naturaleza muy diversa». Rodríguez de las Heras (2025, p. 20).

[87] EU Commission (4 de abril 2021).

[88] Renda *et al*. (abril 2021). Este estudio comparativo vincula explícitamente la cadena de valor con los riesgos y las posibles vulneraciones de derechos humanos, y distingue distintos contextos: «[…] la IA se ha integrado en sistemas más articulados que comprenden objetos inteligentes y una mejor conectividad, lo que plantea nuevos desafíos para la seguridad. Estos riesgos pueden abordarse y mitigarse de diversas maneras. En contextos de empresa a empresa (B2B), las disposiciones contractuales sobre responsabilidad permiten una asignación óptima de responsabilidad a lo largo de la cadena de valor. En contextos de empresa a consumidor (B2C) y de gobierno a ciudadano (G2C), las personas carecen de medios adecuados de acción y reparación, a menudo porque desconocen las prácticas (por ejemplo, las sugerencias indirectas y la manipulación son difíciles de detectar), carecen de disposiciones legales aplicables (por ejemplo, responsabilidad por daños inmateriales) o existen suficientes conocimientos y procedimientos judiciales ágiles». *Ibid*. p. 22.

Estos documentos responden a la perspectiva creada por el *Libro Blanco de Inteligencia Artificial* (2020), basada en dos pilares explícitos: (i) el marco político del denominado «*ecosistema de excelencia* a lo largo de toda la cadena de valor, partiendo de la investigación y la innovación», (ii) un futuro marco normativo para la inteligencia artificial en Europa que generen un *ecosistema de confianza* exclusivo, basado en el «cumplimiento de las normas de la UE, especialmente las normas de protección de los derechos fundamentales y los derechos de los consumidores».

La cadena de valor de la AI se presenta y es empleada, pues, como un concepto fundamental en una normativa europea que pretende abordar *al mismo tiempo* (i) la gestión de un mercado digital europeo en expansión, (ii) la protección de los derechos de ciudadanos y consumidores, además de los derechos humanos y de los fundamentales, (iii) la implementación de los denominados «valores europeos» respecto a la política económica seguir en la EU, (iv) la implementación de valores éticos para la IA, tal como habían sido ya definidos por los cuatro entregables del grupo de expertos en 2018 y 2019[90], (v) la defensa de la «soberanía digital»[91] en la protección y gestión de datos respecto a la dependencia del mercado global, las agendas corporativas de las grandes corporaciones tecnológicas, y su defensa mediante la política pública e intervención en las relaciones internacionales de los estados no miembros con poder para ello (especialmente, USA, Rusia y China), (vi) la construcción de nuevos instrumentos jurídicos "horizontales" para la promoción de la innovación y el uso de la IA en la industria, los sectores estratégicos (como energía, química y farmacia), y las dimensiones definidas por los catorce espacios de datos comunes europeos. La «horizontalidad» de los instrumentos de regulación solo puede comprenderse si éstos se enmarcan en la línea de la cadena de valor entre todos los agentes que participan en la elaboración e implementación de los programas de IA simbólica, y sobre todo generativa, en los productos y servicios introducidos en el mercado.

---

[89] Cf. Renda (2021, pp. 70-71). A partir de McKinsey (Cheatham, 2019), el estudio destaca los riesgos surgidos durante el ciclo de vida de un sistema de IA y las «*medidas generales en las etapas de la cadena de valor de la IA* (desde la concepción hasta el uso y la monitorización». Destaca cinco etapas: (i) *Conceptualización* (en circunstancias específicas para la implementación, seguimiento y monitorización en circunstancias concretas); (ii) *gestión de datos* (los requisitos de calidad de los datos y la detección automática de anomalías ofrecen la posibilidad de evitar el riesgo de datos incompletos o inexactos); (iii) *desarrollo del modelo* (los datos pueden no ser representativos o tener sesgos o ser discriminatorios; por lo tanto, las directrices para la selección de conjuntos de datos de entrenamiento o las pruebas de algoritmos son ejemplos de medidas de control); (iv) *implementación del modelo* (un sistema de IA no debe comercializarse demasiado rápido, para evitar errores de implementación, y deben aplicarse pilotos, pruebas con expertos, pruebas de modelos y pruebas de usuario para evitar estos errores); (v) *uso del modelo y toma de decisiones* (el entorno tecnológico codependiente es necesario para el correcto funcionamiento del modelo y la infraestructura de IA, así como para su alimentación de datos).

[90] Los siete principios éticos finalmente escogidos fueron: 1) acción y supervisión humanas, 2) solidez técnica y seguridad, 3) gestión de la privacidad y de los datos, 4) transparencia, 5) diversidad, no discriminación y equidad, 6) bienestar ambiental y social, y 7) rendición de cuentas. Estos principios se aplican en tres niveles de abstracción, en distintos contextos y escenarios. El equipo elaboró en este sentido una lista de evaluación de la fiabilidad para adaptar los principios al caso específico, además de ─en entregables separados─ 33 recomendaciones de política pública y una escueta exploración de su implementación en tres sectores específicos: (i) administración pública; (ii) salud; (iii) industria en el Internet de las Cosas. Cf. AI HLEG (2020, 4 abril, 20 junio, 17 julio y 23 de julio).

[91] El concepto de «soberanía de datos» (*data sovereignty*) es definido como «el concepto de que los datos están protegidos por la ley y la jurisdicción del estado de su origen, para garantizar los derechos y obligaciones de protección de datos». EU Commission» (2021, *Impact Assessment* AI Act, p. 91). Hay que distinguir esta versión de la «soberanía digital» del *Self-Sovereign Identity paradigm* propio del W3C. Cf. *supra* nota 72. Cf. Rodríguez-Doncel (2025).

*Hay una cadena de obligaciones[92], una cadena de creación de valor económico mediante la IA, una cadena de suministro, una cadena global, y una cadena de cumplimiento normativo (compliance), de las que se habla de forma equivalente sin que hayan sido definidas ni teorizadas.* Solo se define en el Glosario, inmediatamente después del concepto de soberanía digital, la «cadena de valor de los datos» (*data value chain*), sin indicar tampoco qué parámetros van a tenerse en cuenta y cómo pueden ser medidos y comparados a lo largo de la cadena de producción descendente de bienes y servicios.[93]

Hay otras tensiones que cabe señalar. En el Reglamento no solamente aflora la competición y el conflicto abierto (ya no latente) con las grandes corporaciones tecnológicas norteamericanas, sino también la tensión entre la soberanía de los estados y la centralización de poder y de gestión europea. Los principios subyacentes en el Reglamento apuntan en esta dirección: (i) *Subsidiariedad* (reforzar las competencias de la UE frente a los Estados nacionales), (ii) *proporcionalidad* (equilibrar las posiciones de la UE y los Estados nacionales; y los diferentes intereses que cumplen las normas), (iii) *complementariedad* (armonización jurídica), (iv) *seguridad jurídica* (certeza de las disposiciones y normas legales).[94]

Creo que el conflicto con los denominados «gobernantes» de la cadena de valor» emerge en la redacción final del Reglamento, que especifica una serie de sanciones fuertes en la línea ejecutiva vertical del estado de derecho. Hay un predominio y una preocupación evidentes por el desarrollo de corporaciones europeas que puedan hacer frente a las que dominan el mercado global. Los Informes Draghi (tanto el de septiembre de 2024 como su revisión en septiembre de 2025) insisten en la suavización de los requisitos exigidos por los Reglamentos de protección de datos y de IA para promover la competencia, siguiendo el esquema de Porter sobre la creación de ventajas competitivas.[95] El primero ha constituido la base para elaborar la presente estrategia de la UE.

Pese a las proclamaciones que se repiten, la crítica ha señalado el desequilibrio normativo continuado y, en particular, la protección medioambiental. Este es uno de los aspectos más descuidados en el Reglamento, puesto que no existe ninguna autoevaluación al respecto para los desarrolladores e introductores en el mercado. Ugo Pagallo ha puesto de relieve esta anomalía[96], entre otras, como la de haber asumido acríticamente los beneficios de la IA generativa y haber emprendido la revisión excesivamente frecuente de la ley para abordar los avances tecnológicos,[97] como comprobaremos en las secciones siguientes (7 y 8).

---

[92] Ibid. *Impact Assessment*, p. 49. « Obligaciones horizontales vinculantes para todos los actores de la cadena de valor: a. Proveedores de sistemas de IA de alto riesgo [según la Opción 1] + (re)evaluación de conformidad, notificación de riesgos/infracciones, etc. b. Usuarios de sistemas de IA de alto riesgo (supervisión humana, monitorización, documentación mínima)».

[93] «Concepto subyacente que describe la idea de que los activos de datos pueden ser producidos por actores privados o por autoridades públicas e intercambiados en mercados eficientes, como materias primas y componentes industriales (o puestos a disposición para su reutilización como bienes públicos) a lo largo de todo el ciclo de vida de los conjuntos de datos (captura, conservación, almacenamiento, búsqueda, intercambio, transferencia, análisis y visualización). Estos datos se agregan posteriormente como insumos para la producción de bienes y servicios de valor añadido que, a su vez, pueden utilizarse como insumos en la producción de otros bienes y servicios». Ibid. *Impact Assessment*, *Glossary*, p. 91.

[94] Cf. Considerando 176: «Dado que el objetivo del presente Reglamento, a saber, mejorar el funcionamiento del mercado interior y promover la adopción de una IA centrada en el ser humano y fiable, garantizando al mismo tiempo un elevado nivel de protección de la salud, la seguridad y los derechos fundamentales consagrados en la Carta, incluidos la democracia, el Estado de Derecho y la protección del medio ambiente, frente a los efectos perjudiciales de los sistemas de IA en la Unión, así como brindar apoyo a la innovación, no pueden ser alcanzados de manera suficiente por los Estados miembros, sino que, debido a sus dimensiones y efectos, pueden lograrse mejor a escala de la Unión, esta puede adoptar medidas, de acuerdo con el principio de subsidiariedad establecido en el artículo 5 del TUE. De conformidad con el principio de proporcionalidad establecido en el mismo artículo, el presente Reglamento no excede de lo necesario para alcanzar dicho objetivo». *Ibid.* pp. 43-44.

[95] Cf. Draghi (2024, 2025).

### 7. La reciente estrategia europea para el fomento de la innovación y la competencia

El esfuerzo de construir un mapa basado en las múltiples deficiencias y riesgos del desarrollo industrial europeo es encomiable. Ello ha permitido identificar: (i) riesgos de la cadena de suministros (incluida la seguridad energética); (ii) riesgos para la seguridad de las infraestructuras; (iii) riesgos deficitarios respecto al talento; (iv) riesgos de dependencia económica.

En el *Plan de Acción «Continente de IA»* (abril, 2025), se especifican cinco pilares para el incentivo de la competencia: (i) ampliar la infraestructura pública mediante el establecimiento de «gigafactorías eficientes en el uso de los recursos, que integren una capacidad informática masiva en los centros de datos»; (ii) adoptar nuevas medidas «para garantizar un mayor acceso a datos de alta calidad para quienes innoven en materia de IA»; (iii) estimular «un mayor desarrollo de los algoritmos de IA y aprovechar su adopción en los sectores estratégicos de la UE»; (iv) reforzar las capacidades en IA mediante la formación, retención y atracción de talentos; (v) «facilitar el cumplimiento del Reglamento de Inteligencia Artificial, en particular para los innovadores más pequeños».

La recepción de este planteamiento ha sido ambivalente. Se ha tendido a acogerlo de forma prudente, aunque ha habido incluso quien lo ha calificado de tímido, abogando por un principio de «armonización competitiva» más decidida para acabar con la fragmentación normativa.[98] Pero se ha señalado asimismo que el principio que había guiado la política económica hasta ahora, «tan abierta como sea posible, tan autónoma como se necesite»[99], ha sido decantado hacia su converso: «tan autónoma como sea posible, tan abierta como se necesite».[100] Otra vez, ha habido interpretaciones positivas, pues al fin, «quizás la desventaja europea no sea una crisis, sino una condición para la transformación», i.e. puede significar «una catálisis para la autonomía transformadora».[101] Y también negativas, puesto que, según algunos autores, lo que subyace a esta estrategia sería una política que «prioriza la independencia jurisdiccional sobre la soberanía ciudadana, que adopta una lógica de carrera global de IA y que descuida en gran medida su impacto en las personas fuera de la UE».[102]

Desde un punto de vista técnico de la cadena de valor en IA generativa, se ha señalado que «el *AI Continent Action Plan* representa un esfuerzo ambicioso y necesario, pero presenta un enfoque desequilibrado que no aprovecha óptimamente las ventajas competitivas europeas en la cadena de valor de la IA».[103] El programa RAISE, dirigido a la incentivación y promoción de la IA en la ciencia básica y aplicada, busca justamente esta incentivación, la conexión con las empresas y, al mismo tiempo, el impulso del intercambio y la interconexión en ciencia abierta del conocimiento basado en el uso de las técnicas de IA. Se busca armonizar el conocimiento desagregado. Esta política se dirige a los Estados miembros de la UE, siguiendo los cuatro ámbitos prioritarios de desarrollo definidos en la *Recomendación del Consejo de 24 de junio de 2025 sobre los objetivos de actuación del Espacio Europeo de Investigación* [EEI] *para 2025-2027*: (i) fortalecer un mercado interior para el conocimiento realmente funcional; (ii) abordar conjuntamente la transición ecológica y la transformación digital y otros retos con repercusiones en la sociedad, y aumentar la participación de la sociedad en el EEI; (iii) mejorar el acceso a la excelencia en materia de investigación e innovación y mejorar las interconexiones entre los ecosistemas de innovación en toda la Unión; (iv) promover inversiones y reformas concertadas en investigación e innovación.[104] Sin embargo, resulta claro que, además de mencionar como ámbitos prioritarios a los retos ecológicos y sociales (como el de aumentar la participación ciudadana), el primer objetivo es el del desarrollo de la innovación y su aplicación práctica para crear valor en el mercado o «valor añadido europeo»,[105] i.e. *incidir en la cadena de valor*:

> Estas políticas estructurales y acciones se centran en i) *aportar un claro valor añadido en el ámbito nacional y europeo* [énfasis añadido]; ii) generar efectos mediante la producción de resultados concretos y resultados tangibles en un plazo de tres años; iii) basarse en la creación conjunta constante entre los Estados miembros, la Comisión y, cuando corresponda, los países asociados y las partes interesadas; iv) ser autónomas con una actividad principal; y v) aplicarse de forma voluntaria y en geometría variable, lo que permitirá flexibilidad respecto a la medida en que pueden participar los países. Para que se apliquen, las políticas estructurales y las acciones del EEI deben garantizar la participación de al menos la mitad de los Estados miembros.[106]

---

[96] Pagallo (2025b, p.16 ): «El resultado final podría ser que los investigadores tengan que afrontar más cargas legales y responsabilidades ambientales por sus proyectos que las empresas de peso pesado por sus sistemas de IA que no sean de alto riesgo en el mercado: un monstruo legal». Pagallo (2025b, p.10).
[97] Cf. Pagallo (2025a).
[98] Cf. Bauer (2025).
[99] Cf. Schmidt y Seidl (2023).
[100] Cf. Mariotti (2024).
[101] Cf. Koch (2025).
[102] Cf. Mügge (2024).
[103] Cf. Arnal (2025).

La preocupación proviene principalmente del comportamiento y recepción de las empresas, *start ups* y *pymes*. Pero también hay que señalar cierta falta de realismo. Ha habido críticas fundadas al respecto:

> El plan presenta desequilibrios preocupantes y deja sin abordar elementos fundamentales. La apuesta masiva por gigafactorías amenaza con un consumo masivo de recursos financieros para competir en el eslabón donde Europa tiene menor ventaja competitiva. Con 20.000 millones de euros para cinco instalaciones, Europa está intentando replicar el modelo estadounidense de "escala bruta" cuando llega tarde y con presupuestos insuficientes comparados con la competencia global. La escala real del desafío se evidencia al comparar con la competencia: Meta, una sola empresa, ya posee más de 600.000 H100 equivalentes y planea superar 1,3 millones de GPU para finales de 2025, con una inversión de 60.000-65.000 millones de dólares sólo este año. En contraste, las cinco gigafactorías europeas sumarían ~500.000 chips de IA con 20.000 millones de euros total. Europa está intentando competir a nivel continental con lo que una empresa estadounidense ejecuta individualmente y llegando tarde: mientras Meta ya opera con esta escala, las gigafactorías europeas no estarán operativas hasta 2028.[107]

---

[104] Cf. EU Commission (2025, octubre, junio).

[105] Algunos Informes previos resultan meridianamente claros respecto a los objetivos fundamentales que se persiguen. Cf. la primera recomendación del Informe Heitor y Ferguson (2024): «Impulsar un enfoque pangubernamental [*a whole-of-government approach*], alineando una política transformadora de investigación e innovación con la agenda estratégica de la UE y las recientes recomendaciones políticas de alto nivel: a. Impulsar la competitividad de Europa y su posición en la carrera hacia una economía limpia y digital iniciando una nueva era de invención e ingenio. Esto requiere situar la investigación y la innovación, la ciencia y la tecnología, en el centro de nuestra economía; b. Impulsar la integración del mercado europeo y promover la "quinta libertad", que debe abarcar la libre circulación de la investigación, la innovación, el conocimiento y la educación (Letta). Esto requiere reforzar el atractivo de Europa para el talento y las inversiones en I+D+i; c. Implementar un sólido marco europeo de Investigación, Desarrollo e Innovación (I+D+i) *para impulsar el valor añadido europeo y contribuir al establecimiento de una sólida «Unión de la Investigación y la Innovación»* (Draghi) [énfasis añadido].Esto requiere fortalecer el Programa Marco en todo el espectro de la investigación, la tecnología y la innovación.»

[106] Cf. EU Commission (2025, octubre, RAISE, p. 9).

[107] Cf. Arnal (2025). Arnal efectúa un anàlisis equilibrado de los puntós débiles y fuertes de la propuesta.

Por mi parte, me centraré en la perspectiva ética y jurídica. Los documentos programáticos de la EU para hacer frente a la posición dominante de los gigantes tecnológicos —la «*Brújula*» y el «*Continente*» — inclinan la balanza de la cadena de valor hacia la construcción de un mercado digital europeo mediante *gigafactorías, factorías*, y *antenas* que asuman una contraposición dominante, presten servicio a las necesidades de la industria, y aseguren la autonomía y la denominada «soberanía digital» europea. Por lo tanto, interpretan la CVIA casi exclusivamente desde el punto de vista económico (y político) de la competitividad, asumiendo que la construcción de la pirámide de riesgos, la redacción del artículo 25 del la *Ley de IA* y la elaboración de *Buenas Prácticas* es suficiente para lograr *al mismo tiempo* los objetivos éticos y jurídicos de los derechos civiles y humanos.

La impresión que producen estas propuestas es que intentan recoger el guante de la competencia siguiendo la misma estrategia de dominación del mercado que sus rivales, i.e. sitúan la CVIA casi exclusivamente en la esfera de una gobernanza económica y política, pero sin respetar la flexibilidad de los modelos jurídicos del derecho norteamericano y del derecho común (*vid. supra*, sección 2). El Índice AI 2025 de Stanford señala que en 2024 se produjeron 40 nuevos modelos de IA generativa en UA (por 15 en China, y solamente 3 en Europa).[108] La propia UE confirma estos datos.[109] En los últimos veinte años, Europa ha descendido en número de modelos, patentes y publicaciones.[110] Respecto a la investigación en IA, es de notar que China lidera ya en número de publicaciones.[111]

Es preocupante la polarización y que el desarrollo de los mercados de IA se decante tan claramente del lado norteamericano, pero lo es aún más la réplica europea de sus estrategias de mercado en condiciones adversas, máxime cuando hay indicios de que una cuota significativa de la inversión que se realiza en las empresas europeas revierte en realidad al capital de las norteamericanas, y que el capital humano sigue un patrón parecido.[112] Volveremos sobre ello más adelante (sección 8), puesto que el incentivo de los mecanismos de coregulacion al mismo tiempo que la mejora de los procedimientos de protección y de resolución de conflictos, incluyendo especialmente los procedimientos judiciales, puede consolidar y robustecer las vías de gobernanza jurídica de la inteligencia artificial. I.e., constituiría una forma de diseñar cadenas normativas de valor que puede compadecerse con la estrictamente económica.

---

[108] Cf. Gil y Perrault, HAI Stanford (2025, p. 3). Los números son también significativos: «En 2024, la inversión privada estadounidense en IA ascendió a 109.100 millones de dólares, casi 12 veces los 9.300 millones de dólares de China y 24 veces los 4.500 millones de dólares del Reino Unido. La IA generativa experimentó un impulso especialmente fuerte, atrayendo 33.900 millones de dólares a nivel mundial en inversión privada, un aumento del 18,7 % con respecto a 2023. El uso de la IA en las empresas también se está acelerando: el 78 % de las organizaciones informaron haber utilizado IA en 2024, frente al 55 % del año anterior».

[109] Cf. EU Commission (2025, octubre, RAISE, p. 1) : «Los investigadores europeos fueron de los primeros en integrar la IA en su trabajo y, hasta 2017, lideraban el número de publicaciones científicas que utilizaban aplicaciones de IA. Sin embargo, *desde entonces China y los Estados Unidos han alcanzado y superado a la UE*, y ahora el líder a escala mundial es China5. La cuota mundial de la UE en la capacidad computacional de la IA es inferior al 5 %, frente al 75 % en el caso de los Estados Unidos y el 15 % en el de China. Europa sigue siendo un importante centro de investigación básica en materia de IA, que refleja la activa comunidad investigadora en materia de IA del continente. Sin embargo, la UE tiene una pequeña cuota mundial de agentes de IA (6 %) en comparación con los Estados Unidos y China, y una cuota aún menor de patentes de IA (3 %).»

[110] Vid. asimismo, Heitor, Ferguson *et al*. (2024, p. 15).

[111] Vid. los gráficos comparativos en Bianchini *et al*. (2025, p. 14 y ss.).

[112] Cf. McHugh (2025) refiere que aproximadamente el 80 % de las inversiones en IA generativa realizadas en US, Europa e Israel durante 2023 and 2024 (56.000 millones de dólares) siguieron este camino. Dos tercios de la financiación (37.000 millones de dólares) se invirtieron en empresas que desarrollan modelos fundacionales. Cf. AXEL (2025).

Es sorprendente, además, que estos programas reproduzcan el optimismo respecto a la inteligencia artificial general (IAG) a través de la estrategia de la IA generativa que muestran los documentos de Open AI, Palantir o Microsoft,[113] para los que, en aras a alcanzar ese objetivo, no les duelen prendas: la IAG se sitúa como una prioridad que justifica el crecimiento sin límite de la propia corporación. Habría que diferenciar la utilidad de los modelos de LLM con relación a sus condiciones de producción y usabilidad. Respecto a las primeras, las múltiples demandas basadas en la violación de propiedad intelectual han advertido sobre los límites de la apropiación indiscriminada de datos. Respecto a la segunda cuestión, Nicholas Taleb ha usado el símil de *self-liking lollipop* —otra versión de la coleta del barón de Münchhausen— para referirse a la progresiva autoreferencia de la alimentación y producción de resultados de ChatGPT.[114] No se trata, en efecto, de la gestión de conocimiento diseminado en la red, sino de su representación en modelos probabilísticos. Es un instrumento extremadamente útil, con la capacidad de transformación del mercado, una gran incidencia en la gestión de empresas, el mercado laboral, y la vida cotidiana de los ciudadanos, pero de ahí no se sigue que su inmensa potencia de cálculo pueda convertirse y generalizarse en objeto de política pública para la innovación *per se*, y menos para la generación de servicios. Resulta cada vez más clara la necesidad de intervención humana en sus procesos de desarrollo e hibridación con modelos simbólicos y semánticos (vid. *supra* Sección 9). Conviene recordarlo porque el plan del *Continente* se relaciona directamente con esta capacidad.[115]

La innovación de las grandes corporaciones norteamericanas se basa también en una fuerte competición por los segmentos de mercado. Sustituirla por una cooperación mutua coordinada y planificada, como plantea la *Brújula* y su *Anexo*, financiada mediante políticas públicas y legitimada por la autonomía del principio de soberanía digital puede significar en realidad *priorizar la construcción política frente a la social*. Esto deja en el aire la relación estrecha que existe a nivel micro —i.e., en las transacciones reales— entre los mecanismos de negociación y compensación que entraña el principio de responsabilidad contractual y extracontractual, y su efecto económico y social.

Priorizar la autonomía e insistir en la defensa de los valores europeos y la reducción de la dependencia como estrategia genérica (por no hablar del gasto militar y en ciberseguridad) debería entrañar una atención mayor y complementaria a la protección al consumidor, y a la implementación de derechos en situaciones concretas —vulnerabilidad, discapacitación, educación, etc. Sin embargo, a pesar de las declaraciones y proliferación de normativa al respecto, ello no se ha producido y no se ha seguido con la misma atención. En todo caso se presupone como un punto ya

---

[113] La perspectiva que adopta la UE en el Plan de Acción «Continente de IA» (2025, p. 7) es política, y epistemológicamente discutible: «En los dos últimos años, los modelos de IA se han vuelto cada vez más complejos, pasando del procesamiento de textos al razonamiento, las capacidades multimodales y el comportamiento de agente. Esta tendencia continuará, y se espera que la próxima generación de modelos fronterizos de IA desencadene un salto en las capacidades, hacia la inteligencia artificial general capaz de realizar tareas muy complejas y diversas, igualando las capacidades humanas.» Compárese con el estudio de Microsoft (Bubeck, 2023), que se sitúa expresamente en la vía de alcanzar la IAG.
[114] https://x.com/nntaleb/status/1660304315217584128?s=51
[115] «Las gigafactorías de IA serán instalaciones a gran escala que desarrollarán y entrenarán modelos complejos de IA a una escala sin precedentes, con cientos de billones de parámetros. Integrarán una capacidad informática masiva, que superará los 100 000 procesadores de IA avanzados, teniendo en cuenta al mismo tiempo la capacidad de potencia, así como la energía, la eficiencia hídrica y la circularidad. Estas instalaciones son esenciales para que Europa compita a escala mundial y mantenga su autonomía estratégica en el progreso científico y en sectores industriales críticos.» (*ibid.* p. 8)

adquirido en los valores europeos, compatible con una *desregulación* promovida como «suavización» de requisitos administrativos.

Hay que mencionar los dos documentos más recientes que desarrollan esta línea de reforma de los requisitos de la Ley de IA inicial. En octubre de 2025, la estrategia europea intenta reforzarse mediante un plan general de incentivación de la IA en todos los sectores, denominado *Estrategia de uso de la IA*, con tres objetivos:

> (i) Introducir iniciativas emblemáticas sectoriales para impulsar el uso de la IA en sectores industriales clave de la economía de la UE y en el sector público; (ii) abordar los retos transversales para apoyar una mayor adopción de la IA por parte de las pymes; propiciar una mano de obra preparada para la IA en todos los sectores; aprovechar las capacidades avanzadas de IA; y garantizar la confianza en el mercado; (iii) establecer un mecanismo único de gobernanza para liderar un enfoque sectorial y fomentar un proceso dinámico de colaboración entre las partes interesadas.[116]

El *mecanismo único de gobernanza* consiste en tres ámbitos de entes activos para la coordinación del proceso, el *Comité de IA* (coordinación de los estados), la *Alianza para el uso de la IA* (punto de partida único para partes interesadas destinado a configurar las políticas en materia de IA para sectores estratégicos) y el *Observatorio de IA* (para supervisar las tendencias, elaborar indicadores y facilitar el diálogo). Uno de los sectores relevantes —junto con medicina, farmacia, robótica, energía, educación tecnológica, cultura, entre otros— es la administración pública.[117]

Este plan de la *Estrategia de Uso de la IA* (2025) concluye:

> La Estrategia de Uso de la Inteligencia Artificial está diseñada para ayudar a las industrias y al sector público a comprender mejor lo que esta puede hacer, dónde es eficaz y cómo puede aportar ventajas competitivas. Anima a las organizaciones a dar más importancia a la IA en sus esfuerzos de resolución de problemas. Al proponer políticas transversales y sectoriales, la Estrategia ofrece un modelo para apoyar el despliegue y la ampliación de las soluciones de IA pertinentes. Al establecer un mecanismo único de gobernanza, la Estrategia fomenta el diálogo entre los responsables políticos y las diferentes comunidades sectoriales. Al conectar y reforzar los instrumentos relacionados con la IA, sirve de plan rector para la plena adopción e integración de la IA en los sectores estratégicos de la UE, lo que conduce a reforzar el «Continente de IA».[118]

El 19 de noviembre del mismo año, la Comisión publica la propuesta para un nuevo Reglamento *Ómnibus digital sobre la IA* (con Anexos) estableciendo la consulta y participación de las industrias en todos los sectores y espacios estratégicos. Ha sido elaborado para implementar los documentos programáticos y reformar la Ley de IA (2014) en los puntos que deben adecuarse a la nueva estrategia. Esta enmienda se produce *antes* de que el cuerpo de la Ley haya entrado en vigor (el 2 de agosto de 2026 y 2027). Denota, pues, la urgencia de modificarla, puesto que las consultas ya realizadas «revelan desafíos de implementación que podrían poner en peligro la entrada en vigor efectiva de disposiciones clave de la Ley de IA […] La Comisión está adoptando medidas adicionales para facilitar el cumplimiento de la Ley de IA y abordar las preocupaciones planteadas por las partes interesadas.».[119]

---

[116] EU Commission (octubre 2025, p. 2).
[117] La Comisión «creará un conjunto de herramientas de IA dedicado a las administraciones públicas (en particular el poder judicial que cuente con un repositorio compartido de herramientas y soluciones prácticas, de código abierto y reutilizables, para apoyar la interoperabilidad de la IA. Este conjunto de herramientas también incluirá las soluciones de IA previstas en la Hoja de ruta para un acceso lícito y efectivo a los datos por parte de las autoridades policiales y judiciales. Además, se pondrá en marcha *la Public Sector AI & Interoperability Readiness Pathway (PAIR Pathway)* para proporcionar ejemplos prácticos paso a paso dentro del itinerario de los usuarios que ayudará a las administraciones a desarrollar servicios adaptados a sus necesidades específicas.» [117] EU Commission (octubre 2025, p.14).
[118] EU Commission (octubre 2025, p. 23).

Es un conjunto extenso y complejo de enmiendas del articulado anterior para promover la consistencia de las normas de la Ley de IA, facilitar su cumplimiento, simplificando los trámites, pero especificando que se mantendrá «el mismo nivel de escrutinio».[120] Puede compararse con el tipo más horizontal de consultas efectuadas, por ejemplo, en Canadá, para comprobar la diferencia de posición respecto a las empresas, las cuales deben seguir demostrando que cumplen con los requerimientos, aunque estos se simplifiquen.[121] Una de las medidas a desarrollar son las «Directrices sobre la aplicación práctica de las normas sobre responsabilidades [*responsibilities*] a lo largo de la cadena de valor de la IA».[122] Pero, ¿en qué consiste esta cadena de valor de la IA?

## 8. La cadena ética y jurídica de valor

Al mismo tiempo que se publicaba el *Informe del Continente Europeo* en febrero de 2025, la Comisión retiraba la *Propuesta de la Directiva sobre responsabilidad en materia de IA* que ya se había presentado como proposición. Esto es inusual: no había razones jurídicas para ello.[123] Las razones, en efecto, eran políticas. El eurodiputado Axel Voss, ponente de la Directiva, condenó la decisión de la Comisión, afirmando que ésta había optado decididamente por «la incertidumbre jurídica, los desequilibrios de poder corporativo y un enfoque del Salvaje Oeste que sólo beneficia a las grandes tecnológicas».[124] No le faltaba razón. No sólo se retiró esta proposición, sino treinta y seis más (algunas asimismo importantes, sobre protección de derechos, transparencia y antidiscriminación).[125] Fueron consideradas «redundantes». En el caso de la responsabilidad extracontractual, la Propuesta daba soluciones para una responsabilidad *ex post*, y no *ex ante*,[126] como la Ley de IA. En su Exposición de Motivos, la Propuesta especificaba:

> Las normas nacionales en vigor en materia de responsabilidad civil, particularmente las que se basan en la culpa, no son adecuadas para tramitar las denuncias de responsabilidad civil por daños causados por productos y servicios en los que se recurre a la IA. Con arreglo a dichas normas, las víctimas deben demostrar que ha habido una acción u omisión ilícita por parte de una persona que ha causado el daño. Las características específicas de la IA, incluidas su complejidad, su autonomía y su opacidad (el denominado efecto de «caja negra»), pueden dificultar o hacer excesivamente costoso para las víctimas determinar cuál es la persona responsable y probar que se cumplen los requisitos para una demanda de responsabilidad civil admisible. En particular, al reclamar una indemnización, las víctimas podrían tener que soportar unos costes iniciales muy elevados y enfrentarse a procedimientos judiciales mucho. más largos, en comparación con los casos sin relación alguna con la inteligencia artificial. Por lo tanto, las víctimas pueden verse disuadidas de intentar siquiera obtener una indemnización.[127]

Especialmente la carga de la prueba de la culpa y la causalidad pueden resultar difíciles de probar. En los tribunales norteamericanos, la doctrina del «uso legítimo» [*fair use*] ha legitimado el entreno de grandes modelos de lenguaje como CLAUDE.[128] Y en algunos ordenamientos europeos se exige explicitar la causación del daño, i.e. la relación entre la causa alegada y el efecto producido, cosa que, en los LLMs, debido a su opacidad, acaba siendo imposible. Así que en la Directiva se introducía una presunción de responsabilidad refutable (*juris tantum*), excepto cuando el demandado demostrase que el demandante podía acceder razonablemente a pruebas y conocimientos especializados suficientes para demostrar el nexo causal (en el caso de los sistemas de IA de alto riesgo).[129] La evaluación de impacto que acompañaba a la Propuesta hacía también hincapié en los efectos de la seguridad jurídica y la reducción de la incertidumbre sobre la cadena de valor:

> La falta de confianza y la consiguiente menor adopción de productos y servicios basados en IA probablemente afectarán a todas las empresas involucradas en el despliegue de la IA, generando pérdidas de oportunidades de mercado. *Esto incluye también a las entidades de la cadena de valor de la IA, como investigadores, desarrolladores y proveedores de tecnologías o equipos relacionados (por ejemplo, proveedores de servicios en la nube), quienes probablemente se enfrentarán a una menor demanda* [énfasis añadido].[130]

---

[119] EU Commission (novembre, 2025, *Explanatory Memorandum*, p. 1-2).
[120] EU Commission, *Omnibus,* Considerando 7, p. 13: «Para garantizar la coherencia, evitar duplicaciones y minimizar las cargas administrativas en relación con el procedimiento de designación de organismos notificados en virtud del Reglamento (UE) 2024/1689, *manteniendo el mismo nivel de escrutinio*, debe estar disponible una solicitud única y un procedimiento de evaluación único para los nuevos organismos de evaluación de la conformidad y los organismos notificados […]».
[121] Government of Canada. Competition-Bureau (enero, 2025).
[122] *Omnibus*, *ibid*. p. 3.
[123] Cf. De Capitani (2025, febrero). Cf. también Du Brocard (2025, febrero).
[124] Cf. Datta (2025, febrero).
[125] Véase la lista completa en De Capitani (2025), op. cit.
[126] Cf. Nannini (2025) ofrece una explicación y un anàlisis detallados de la situación.
[127] EU Commission (2022, septiembre, p. 1-2).

Así, la evaluación concluía que:

> La consecuencia práctica de esta situación es injusta y económicamente ineficiente. Por un lado, quienes asumen riesgos y se benefician de la integración de la IA en sus actividades podrán repercutir en las víctimas los costes asociados a su uso. Por otro lado, las víctimas —ya sean personas físicas o jurídicas— se verán sobrecargadas con costes que no pueden evitar. El marco de responsabilidad civil, por tanto, no cumpliría su función de prevenir conductas indeseadas y garantizar una asignación eficiente de los costes asociados a los riesgos.[131]

La Propuesta se basaba en la posibilidad de efectuar demandas de responsabilidad civil extracontractual[132], coadyuvada por mecanismos que las facilitasen. Era un instrumento ya existente, suficiente para armonizar la finalidad de protección y compensación de daños para todas las

---

[128] Por ejemplo, en la demanda contra ANTHROPIC presentada por algunos escritores por infringir los derechos de autor de sus obras para entrenar a CLAUDE, el juez de distrito William Alson ha preservado recientemente el derecho de uso de la compañía con la siguiente argumentación: «En resumen, el propósito y la naturaleza del uso de obras protegidas por derechos de autor para entrenar a LLMs era en esencia transformador. Al igual que cualquier lector que aspira a ser escritor, el LLM de Anthropic se alimentaba con obras no para replicarlas o suplantarlas, sino para innovar y crear algo diferente. Si este proceso de formación requería razonablemente realizar copias dentro del LLM o de otra manera, dichas copias se utilizaban con fines transformadores. El primer factor favorece el uso legítimo [*fair use*] de las copias de entreno.» El juez también especifica que el argumento no cubre las copias pirateadas que también han sido empleadas para el entreno. Cf. USDC (2025, junio, p.13-14).
[129] Cf. EU Commission (2022, septiembre) art. 4 y art. 5).
[130] Cf. EU Commission (2022, septiembre, Impact Ass., p. 17). También: «Incrementar la confianza social beneficiaría a todas las empresas de la cadena de valor de la IA, ya que fortalecer la confianza ciudadana contribuiría a una adopción más rápida de la IA y crearía una ventaja competitiva para las empresas europeas del sector. Debido al efecto incentivador de las normas de responsabilidad, evitar lagunas legales también beneficiaría indirectamente a toda la ciudadanía mediante una mayor protección de la salud y la seguridad (artículo 114, apartado 3, del TFUE) y la eliminación de fuentes de riesgo para la salud (artículo 168, apartado 1, del TFUE).» (*ibid. p.* 25). Incorporado al texto final de la Propuesta: «Contribuirá a un régimen de responsabilidad civil eficiente, adaptado a las especificidades de la IA, en el que las demandas fundamentadas de indemnización por daños y perjuicios sean estimadas. *El aumento de la confianza social también beneficiaría a todas las empresas de la cadena de valor de la IA, ya que el aumento de la confianza de los ciudadanos contribuirá a una adopción más rápida de la IA*» [énfasis añadido]. EU Commission (2022, septiembre, p. 5).
[131] EU Commission (2022, septiembre, Impact Ass., p. 28).
[132] Art. 2.5: «Demanda por daños y perjuicios»: una demanda de responsabilidad civil extracontractual subjetiva (basada en la culpa) por la que se solicita una indemnización por los daños y perjuicios causados por una información de salida de un sistema de IA o por la no producción por parte de dicho sistema de una información de salida que debería haber producido.»

jurisdicciones de los países de la EU y evitar múltiples interpretaciones distintas según su objeto en campos tan dispares como las finanzas, la propiedad intelectual, el diseño industrial, la seguridad viaria o el propio procedimiento judicial. Aclaraba cuestiones relativas a la carga de la prueba y a las condiciones de revelar información confidencial sobre los algoritmos que constituían —y siguen constituyendo— problemas recurrentes.[133] Además, facilitaba una mejor asignación de costes a lo largo de la cadena de valor para las compañías aseguradoras.[134]

Paradójicamente, la retirada de esta Propuesta ha significado volver al estado anterior. Como ha señalado la crítica, «esta situación ilustra lo que podría denominarse la "*falacia de la fragmentación*", es decir, *la creencia de que menos normas a nivel de la UE implican necesariamente una menor carga regulatoria*» [énfasis añadido].[135]

Cuatro meses antes de su retirada, en septiembre de 2024, La Comisión de Asuntos Jurídicos (JURI) del Parlamento Europeo solicitó una evaluación de impacto complementaria de la propuesta. El nuevo análisis crítico identificó deficiencias clave en la evaluación de impacto de la Comisión Europea, entre las que destacaban una exploración incompleta de las opciones de política regulatoria y un análisis de coste-beneficio reducido, en particular del régimen de responsabilidad objetiva. El nuevo estudio proponía que la AILD ampliase su ámbito de aplicación para incluir sistemas de IA de propósito general y otros sistemas de IA de gran impacto, además del *software*. Asimismo, analizaba un marco de responsabilidad mixta para equilibrar la responsabilidad por culpa y la responsabilidad objetiva. El estudio recomendaba no la regulación mediante una Directiva —que requiere la aprobación de los parlamentos nacionales— sino directamente mediante un Reglamento sobre la responsabilidad del *software*, para evitar la fragmentación del mercado y aportar claridad a las interpretaciones. Planteaba asimismo un retorno a la propuesta anterior de responsabilidad solidaria (*joint liability*). Así:

> El principal dilema en torno a la responsabilidad solidaria radica en la distribución de la responsabilidad a lo largo de la cadena de valor de la IA. El artículo 25 de la Ley de IA, incorporado en el último estadio de la ley, define ciertas situaciones en las que el implementador [*deployer*][136] se convierte en el nuevo proveedor, asumiendo la responsabilidad principal del cumplimiento de la Ley de IA. Una disposición similar no figura ni en la Directiva de Protección de Datos (DPD) revisada ni en la Directiva de Inteligencia Artificial (DIA). El problema fundamental de la responsabilidad a lo largo de la cadena de valor puede formularse, en principio, de la siguiente manera: para garantizar una indemnización efectiva, el último actor de la cadena —aquel que se enfrenta a la parte perjudicada, a menudo un implementador (el actor final de la cadena)— debería ser considerado responsable si se cumplen las condiciones de responsabilidad previstas en la legislación nacional. Dicha entidad o persona suele ser el punto de contacto más accesible para la persona perjudicada. Sin embargo, otras entidades anteriores en la cadena también deberían ser responsables, bajo ciertas condiciones definidas principalmente por la legislación nacional: primero, porque el último actor podría ser insolvente; y segundo, porque la culpa real podría recaer en otras entidades de la cadena de valor.[137]

---

[133] Véase, por ejemplo, la Opinión del Abogado General de la EU Richard de la Tour (2024, septiembre), a petición del Magistrado de la Ciudad de Viena, sobre el significado del concepto de «información significativa sobre la lógica implicada». Se trata de un caso de daños, por un lado, y de protección de los conocimientos técnicos de información comercial, por otro. La solicitud se realizó en el contexto de un conflicto entre CK y el *Magistrat der Stadt Wien* (Ayuntamiento de Viena, Austria) relativa a una solicitud de ejecución de una orden judicial que exigía a una empresa de evaluación crediticia proporcionar a CK información significativa sobre la lógica involucrada en la elaboración de perfiles relacionados con sus datos personales. A CK le fue denegada la renovación de un contrato de telefonía móvil con una cuota mensual de 10 euros, alegando que no tenía suficiente solvencia económica. Esta supuesta falta de solvencia se corroboró mediante una evaluación crediticia automatizada, empresa especializada en evaluaciones crediticias. CK solicitó a la autoridad austriaca de protección de datos información detallada sobre la lógica aplicada en la elaraciñon de su perfil para la toma automática de decisiones. Dicha autoridad accedió a su solicitud. El Abogado General debía esclarecer qué debe entenderse por «información significativa sobre la lógica implicada» en la toma de decisiones automatizada, en el sentido del artículo 15(1)(h) del Reglamento General de Protección de Datos (RGPD), y cómo debía ponderarse el derecho de acceso a dicha información frente a la protección de los derechos y libertades de otros (secretos comerciales, en este caso). Dudas de este tipo son frecuentes en la interpretación de las leyes sobre tecnología e IA y dificultan la resolución rápida de los casos y la materialización efectiva de los derechos de los ciudadanos.

[134] «Por ejemplo, si el fabricante tiene una reclamación de responsabilidad civil contra el implementador, la aseguradora del fabricante tendría la posibilidad de ejercer un recurso con base en esta reclamación subrogada contractualmente contra el implementador. En última instancia, la aseguradora del implementador sería la que pagaría. Al mismo tiempo, la cobertura de seguro de las personas responsables limitaría los costos económicos a la prima anual del seguro y mantendría bajos los requisitos de entrada al mercado para los fabricantes y operadores de IA, garantizando al mismo tiempo que el daño sufrido por la víctima se indemnizara de forma rápida». EU Commission (2022, septiembre, Impact Ass., p. 55).

[135] Nannini (2025, p. 1851).

Phillip Hacker, el autor de la evaluación de impacto complementaria, sintetiza así su propuesta:

> ***Responsabilidad solidaria a lo largo de la cadena de valor.*** El reparto equitativo de la responsabilidad en la cadena de valor de la IA, especialmente en los sistemas de IA de propósito general, exige un marco de reparación simplificado en la Directiva sobre IA (AILD), para reducir la dependencia de diversas legislaciones nacionales y fomentar un entorno jurídico coherente en la UE. Partiendo de las disposiciones de la Directiva sobre Protección de Datos (PLD) y la Ley de Protección de Datos, tres opciones políticas —la presunción de responsabilidad compartida, el apoyo a las PYME y la protección de las partes posteriores en la cadena de suministro [*downstream parties*][138]— ofrecen distintos mecanismos para garantizar una compensación y una distribución de la responsabilidad justas. Estas opciones pueden combinarse y todas requieren disposiciones vinculantes para evitar que las prácticas contractuales menoscaben estas protecciones.[139]

En mi opinión, aunque no lo califica de ese modo, esto significa que para regular de forma eficiente el sistema de responsabilidad de la IA no es suficiente la aplicación de la cadena de valor como un concepto clave para el análisis de la competitividad tal como ha sido entendida hasta ahora y reflejada en los documentos programáticos de la UE. Resulta necesaria también la construcción de una *cadena jurídica de valor de la IA, en este caso a lo largo de la responsabilidad enlazada de los distintos actores.*

La cadena debería seguir las fases ya identificadas en la cadena de valor, pero especificando los requerimientos normativos (permisiones, obligaciones, prohibiciones) que se establecen en los distintos instrumentos, fuentes, niveles y dimensiones relevantes del estado de derecho.[140]

---

[136] La versión castellana del Reglamento traduce *deployer* por «responsable del despliegue». El Considerando 13 lo define de la siguiente manera: «El concepto de «responsable del despliegue» a que hace referencia el presente Reglamento debe interpretarse como cualquier persona física o jurídica, incluida cualquier autoridad pública, órgano u organismo, que utilice un sistema de IA bajo su propia autoridad, salvo cuando su uso se enmarque en una actividad personal de carácter no profesional. Dependiendo del tipo de sistema de IA, el uso del sistema puede afectar a personas distintas del responsable del despliegue.» (EU Commission (junio, 2024, p. 4).

[137] EPRS (2024, septiembre, p. 33) [Hacker]. Las críticas a la Directiva habían sido ya avanzadas en el análisis efectuado en Hacker (2023).

[138] *Downstream provider*, *downstream system* o *downstream risk* son términos de difícil traducción. En este caso *downstram parties* ha sido traducido por «partes posteriores en la cadena de suministro».

[139] EPRS (2024, septiembre, p. IV [Hacker].

[140] Pueden seguirse las etapas del modelo económico de la cadena de valor para IA ya propuestas, por ejemplo, por Küspert *et al.* (2023), Chardon-Boucaud *et al.* (2024), Arnal (2025), Attard-Frost y Widder (2025), Cuofaro (2025), Wang *et al.*, Wegner (2025), Billones *et al.* (2025). Son distintas entre sí, y más o menos granulares. No vamos a compararlas ahora. Quede para un estudio posterior.

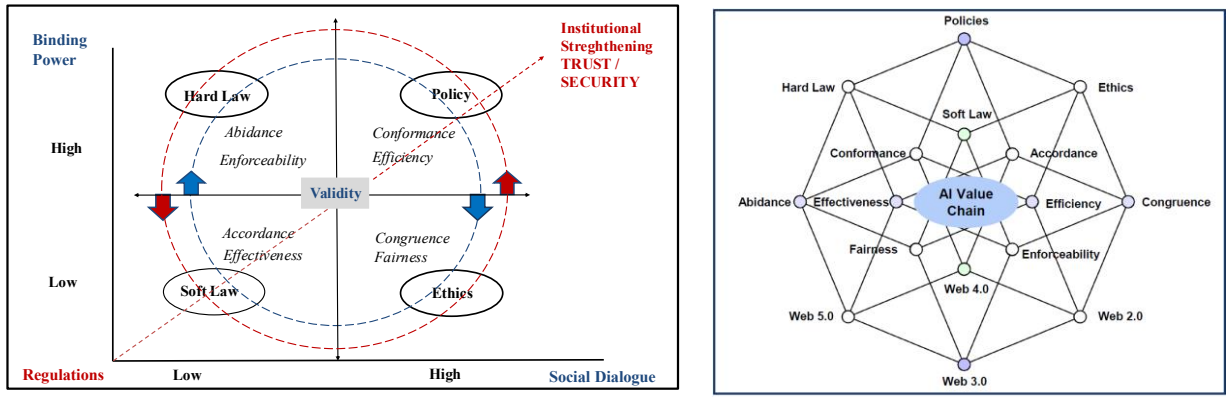

**Fig. 4.1.** Instrumentos de regulación del estado de derecho: *clusters*, atributos y fuerza normativa (grado y tipo de validez).

**Fig. 4.2**. Centralidad de la cadena de valor de la IA en las dimensiones de los ecosistemas jurídicos (compárese con la Fig. 2, *supra*).

Hemos reproducido ya algunas veces a modo de compás o brújula orientativa el cuadrante de la Figura 4.1 relativo a las fuentes del estado de derecho respecto a su valor de *validez* jurídica, situada en el centro de los ejes horizontal y vertical del diagrama.[141] Se trata de una clasificación de los instrumentos jurídicos del estado de derecho en función: (i) del valor que poseen como derecho ejecutable (i.e. coercitivo) o como derecho negociable (i.e. relacional), (ii) de la autoridad de la que emanan las normas (parlamento, tribunales, agencias, sociedades profesionales, etc.), (iii) de los atributos que delimitan su validez jurídica (ejecutividad, eficiencia, efectividad, justicia), (iv) del carácter estatal o social de la relación normativa con los sujetos activos, pasivos (o ambos) de las normas (con distintos roles: ciudadanos, consumidores, trabajadores, empresarios, etc.).

A su vez, la Figura 4.2 reproduce de forma simplificada el diagrama de la Figura. 2, en el que se da volumen y profundidad a los cuatro ámbitos de la clasificación anterior añadiendo las dimensiones en que se sitúan sus contextos tecnológicos de uso (Web 2.0 ─información─, 3.0 ─semántica─, 4.0 ─agentes─, 5.0 ─normativa o axiológica). Se trata de una imagen simplificada de las tecnologías de la web de datos enlazados y de servicios, la web de multiagentes inteligentes, y los sistemas ciberfísicos del denominado internet de las cosas, pero creo que es suficiente para redimensionar la gobernanza multinivel jurídica en los entornos de «inteligencia híbrida» entre humanos y máquinas. Cuando podemos definir quienes son los actores que intervienen en la cadena (diseñadores, desarrolladores, proveedores, implementadores, usuarios, etc.), el resultado no solamente es un modelo de IA o un sistema normativo, sino un ecosistema jurídico implementado en contextos sociales y ámbitos institucionales concretos (fábricas, hospitales, residencias, administraciones, etc.).

En la UE, hasta el momento, parece que se da por supuesto o, a lo sumo, (i) se delega en las pequeñas y medianas empresas la carga de la prueba, i.e., la obligación de probar que «cumplen» con los requisitos establecidos por la ley; (ii) se vuelven a crear más oficinas administrativas para «ayudarles» a cumplir con este objetivo (aparte de las ya existentes por la aplicación de la Ley de IA). En ambos casos, son mecanismos verticales de gobernanza, y no horizontales. «Horizontalidad» se interpreta como interseccionalidad con relación a los distintos espacios de datos y sectores industriales, no como formas de cogobernanza o de autonomía de los productores o empresas (incluyendo las decisiones respecto a su moralidad y comportamiento jurídico).

---

[141] Cf. el proceso de construcción y justificación del modelo en Casanovas, Hashmi, de Koker (2021).

Por otra parte, el rápido desarrollo de nuevos mercados sigue centrando la atención en la actuación e intervención vertical de las autoridades de la competencia. La creación de grandes centros de datos y el crecimiento de las corporaciones tecnológicas ha dado lugar a la identificación de la denominada «cadena de suministro de la infraestructura de IA» (*AI infrastructure supply chain*) (incluyendo los centros de datos y sus recursos de computación). Un estudio pionero de la OECD sobre este tipo de mercados asevera:

> De cara al futuro, las autoridades de la competencia podrían tener que considerar el alcance total de sus facultades políticas y de aplicación de la ley al interactuar con este mercado en rápida evolución. En las primeras etapas, herramientas como la defensa, la disuasión y el control de fusiones pueden desempeñar un papel especialmente importante a la hora de moldear el comportamiento del mercado y guiar el desarrollo del sector. Sin embargo, los responsables políticos también deberán evaluar si las facultades existentes son suficientes para abordar posibles abusos de posición dominante, en particular cuando puedan surgir efectos excluyentes. Esto podría incluir la evaluación de la necesidad de medidas *ex ante*, como los requisitos de interoperabilidad, para complementar la aplicación tradicional. Además, si bien las preocupaciones en torno a la exclusión y el poder de mercado pueden atraer una atención considerable, las autoridades deben mantenerse vigilantes en otras áreas. La escala de la construcción de infraestructura en curso sugiere que riesgos como la manipulación de licitaciones pueden volverse más notorios y, a medida que las tecnologías maduran y se mercantilizan, el potencial de coordinación de precios o comportamiento de cártel también puede aumentar.[142]

Como ha sido anticipado (vid. *supra* sección 5.6), la dimensión ética y jurídica de la CVIA requiere asimismo la coordinación con el cálculo de la cadena genérica de valor. Son aún pertinentes las preguntas formuladas por los investigadores del Ada Lovelace Institute respecto a *la CV de la IA de uso general* (*general-purpose AI, GPAI*) introducida por el Reglamento y que han sido sólo parcialmente resueltas por el Código de Buenas Prácticas: (i) ¿Cómo deberían definirse efectivamente los GPAI?, (ii) ¿Cómo incluir en el cálculo los distintos incentivos de las estrategias de lanzamiento de código abierto?, (iii) ¿Qué instrumentos de regulación se requieren para garantizar la sostenibilidad y gobernanza futura de los modelos GPAI?, (iv) ¿Cómo pueden garantizar los reguladores una capacidad adecuada de supervisión para abordar las implicaciones a nivel social de los modelos GPAI?, (v) ¿Cómo puede la gobernanza de la IA incorporar las opiniones de las personas y comunidades afectadas, de modo que los beneficios de la GPAI se distribuyan equitativamente?[143]

## 9. Lengua y CVIA

Voy a finalizar con dos ideas relativas a la relación entre la cadena de valor de la IA y la lengua. La primera es muy simple, y se refiere a la necesidad de tener un vocabulario estandarizado para la interoperabilidad lingüística en materia de regulación, derecho e inteligencia artificial. Corresponde a lo que en el X Congreso de la Lengua Española ha venido refiriéndose como "lenguaje claro". No parece ser así si se comparan los términos en inglés y en castellano del *Reglamento de Inteligencia Artificial Europeo*. Aparentemente se trata solamente de una traducción lexicográfica de términos, pero en realidad se trata de la validez jurídica de los conceptos que aparecen como tales en la disposición normativa y que pueden inducir a error. La Figura 4 compara la equivalencia del glosario del art. 3 entre la lengua inglesa y española. *Privacy* se traduce por "intimidad" (el término clásico del Código Penal español) y no por "privacidad"; *harmonized standard* por

---

[142] OECD (2025, noviembre, p. 43).
[143] Cf. Küspert, Moes, y Dunlop (2023).

"norma armonizada", y no por "estándar armonizado"; *Law Enforcement Authority (LEA)*, el acrónimo usado en inglés para suavizar lo que significa simplemente "policía", se traduce por "autoridad garante del cumplimiento del derecho", obviando que *enforcement* significa ejecución, cumplimiento forzoso; *regulatory sandbox*, literalmente "parque regulatorio", se traduce por "espacio controlado de pruebas para la IA"[144], *general-purpose AI model* por modelo «de uso general», y *deep-fake* por un término imposible, "ultrasuplantación". Quizás habría que regularizar el uso de los términos del lenguaje normativo en castellano y su equivalencia al inglés, puesto que las redacciones oficiales de los documentos son los que delimitan el significado de los términos jurídicos y su valor. En la práctica profesional, los despachos ─y la Agencia de Protección de Datos[145]─ han regularizado el uso de «privacidad», con un significado más amplio que el de «intimidad».[146]

La segunda idea que me gustaría introducir se refiere a la cultura jurídica y a la tecnología que permite describirla, gestionarla y hacerla evolucionar. Resulta ya común en numerosos estudios de derecho computacional (*computational law*) recurrir a los grandes modelos de lenguaje para el análisis normativo. La Comisión de Asuntos Jurídicos (JURI) del Parlamento Europeo encargó un estudio en profundidad para determinar el alcance de la IA, publicado el 30 de octubre de 2025.[147] Constituye un excelente repaso de las técnicas y metodologías existentes, y apuesta por una perspectiva «híbrida» para «monitorizar» la aplicación de la legislación europea. Las soluciones híbridas, «que combinan el poder de los LLM con el conocimiento generado y comprensible por seres humanos, pueden brindar soluciones apropiadas en algunos dominios, mejorando la capacidad de control, la transparencia y la integración con las aplicaciones existentes».[148] Respecto a la monitorización mediante sistemas de inteligencia artificial, esta es entendida como la necesidad de explicar, i.e., traducir en lengua natural y proporcionar explicaciones jurídicamente relevantes. «Esto significa que deben justificar sus conclusiones citando artículos jurídicos y pruebas específicas de forma que los humanos puedan comprenderlas».[149] Existe a través de todo el documento la tensión entre las lenguas naturales mediante las que opera (y se opera en) el derecho en todas sus fases ─creación, redacción, expresión, interpretación y aplicación de leyes y sentencias─ y su expresión en lenguajes formales (lógicos) y computacionales. El documento advierte de las dificultades encontradas con los sistemas de LLM, pero anima a efectuar un uso controlado de ellos.[150] Pasa también revista a instrumentos ya existentes en la propia Comisión, como THEMIS (*Tool for Harmonised European Monitoring and Information System*)[151] (junto con el proyecto europeo THEMIS5.0)[152] y GPT@EC[153]. Este último es un instrumento de reciente creación

---

[144] Asunción Gómez-Pérez (2023, pp. 91-92) traduce más exactamente la expresión como "espacio controlado de pruebas de regulación".

[145] Cf. Agencia EPD (2025, p. 2) «Nuestra misión es garantizar de manera efectiva y proactiva el derecho fundamental a la protección de datos personales de la ciudadanía, con especial atención a los colectivos vulnerables. Para ello, *apostamos por la privacidad desde el diseño y por defecto como enfoque prioritario* [énfasis añadido]».

[146] Según la Real Academia de la Lengua, 'privacidad' refiere al «ámbito de la vida privada que se tiene derecho a proteger de cualquier intromisión», e 'intimidad' a la «zona espiritual íntima y reserrada de una persona o de un grupo, especialmente de una familia».

[147] Cf. Sartor y Dal Pont (2025).

[148] Sartor y Dal Pont (2025, p. 53).

[149] *Ibid*. p. 53.

[150] *Ibid*. p. 51

[151] https://interoperable-europe.ec.europa.eu/interoperable-europe/themis

[152] *Human-centered Trustworthiness Optimisation in Hybrid Decision Support*, https://www.themis-trust.eu/

[153] Como es definido por la propia Comisión: «La herramienta GPT@EC permitirá al personal mejorar su productividad y creatividad, ayudándolos en tareas como la redacción de textos, el resumen de documentos,

para poner la IA generativa al servicio del personal de la Comisión, basado en la experiencia previa de GPT@JRC (2023).[154]

El Informe de JURI se centra en el paso (y la posible integración o combinación) de los modelos simbólicos a los generativos (fundacionales y LLM) adoptando la perspectiva genérica de los instrumentos y los sistemas normativos, i.e. sin la perspectiva regulatoria propia de su implementación en espacios de datos concretos y en cadenas de valor. Dentro del estado de derecho, adopta el punto de vista estructural de la legislación y la jurisprudencia (*hard law*), ampliando su ámbito al de los *smart contracts* y a los nuevos instrumentos de la denominada *Better Regulation* europea (como, por ejemplo, los espacios controlados de datos, *sandboxes*). Esto describe una parte importante del derecho privado y, especialmente, público, pero sin detenerse en las condiciones para la creación de valor partiendo desde el interior de los procesos de información en las plataformas y flujos de información —*middle-out* e *inside-out*— que es lo que caracteriza a la gobernanza jurídica, i.e. el diseño y emergencia de ecosistemas jurídicos desde el punto de vista empírico en su implementación en contextos determinados.

La IA generativa puede moldearse mediante ingeniería de indicaciones o instrucciones (*prompt engineering*), cadenas de pensamiento (*chain of thought*), modelos conversacionales, medidas de reajuste o *refinamiento* (fine-tuning). Estas técnicas se utilizan en la elaboración de modelos de derecho computacional adaptados al lenguaje jurídico que requieren entrenamiento con una enorme cantidad de datos. Pero no son la única vía. En el campo de los modelos de regulación y los sistemas éticos y/o jurídicos, el modelado a partir de reglas se enfrenta aún a problemas que suelen resolverse *ad hoc* —como el del permiso débil.[155] Es necesario subrayar la importancia de los factores humanos, la cultura, los lenguajes especializados, las anotaciones y el conocimiento de los contextos de uso. Las propuestas de «programación jurídica» tienen en cuenta todo el ciclo y la participación del conocimiento experto en la elaboración de reglas formales.[156] Las técnicas de inyección de conocimiento (*knowledge or semantic injection*) en la entrada de datos o en el propio modelo de lenguaje pueden acotar el modelo a finalidades más precisas y con menos parámetros. Así, después de ALIA,[157] el documento para la estrategia de la IA en España de 2024

---

el desarrollo de código de software, etc., a la vez que protege sus datos. La herramienta cumple con las normas corporativas y los estándares de cumplimiento. Su principal ventaja frente a las herramientas de terceros es su adaptación al contexto de la Comisión, lo que permite procesar la información interna de la Comisión sin compartirla con terceros. GPT@EC se centra en apoyar al personal con una mayor variedad de tareas, proporcionando acceso a varios Grandes Modelos de Lenguaje (LLM), cada uno adecuado para diferentes tareas y niveles de sensibilidad. También ofrece la posibilidad de utilizar los modelos de código abierto y soberanos». https://commission.europa.eu/news-and-media/news/commission-launches-new-general-purpose-ai-tool-gptec-2024-10-22_en

[154] Cf. Para un análisis de su uso, Fernández-Machado *et al.* (2025): «El programa define tres niveles de uso: (i) el N1 implica el uso básico de las capacidades de LLM en las tareas diarias; (ii) el Nivel 2 está dirigido a empleados técnicamente capacitados que utilizan la API para integrar GenAI en herramientas existentes; y (iii) el Nivel 3 cubre a los usuarios avanzados que conectan las capacidades de LLM a fuentes de datos internas y externas».

[155] Vid. *supra* nota 59. Sólo a partir de una perspectiva teórica, i.e. de metalógica, que pueda guiar la formalización es posible ofrecer soluciones para el caso de conflicto entre dos obligaciones. Cf. Governatori y Rotolo (2025).

[156] Cf. Cristani *et al.* (2025).

[157] *Palanca 3: Generar Modelos y Corpus para una infraestructura pública de Modelos de Lenguaje*: "familia de modelos de lenguaje desarrollados bajo criterios de transparencia y confiabilidad, con un aumento de la participación del castellano y las lenguas cooficiales, lo que les diferencia de los modelos dominantes centrados en el inglés Es uno de los principales objetivos estratégicos de este plan. Este modelo de lenguaje se denomina ALIA." Cf. Ministerio de Transformación Tecnológica (2024), p. 22.

apuesta por sistemas más controlados que aumenten el valor añadido y generen ecosistemas (incluyendo el comportamiento sostenido de los usuarios dentro del sistema regulatorio):

> A partir del desarrollo de los modelos fundacionales [...], el desarrollo de modelos especializados (lo que se conoce como *small language models* o SLM), constituyen una palanca para generar un ecosistema empresarial nacional robusto y diverso en este ámbito, que pueda de este modo impulsar servicios de valor añadido basados en los recursos tecnológicos que se apoyan en la lengua, y que favorezca una mejora sostenida de la competitividad a nivel nacional e internacional para las empresas. [...] Estos modelos son un complemento eficiente a los grandes modelos fundacionales por su capacidad para realizar tareas específicas en dominios concretos con menor demanda de recursos computacionales.[158]

Esta estrategia afecta a la concepción de la cadena de valor de la IA, puesto que permite construirla en función de las cinco vertientes o aspectos que han sido definidos anteriormente. Es decir, y esto me parece relevante, la creación de las cadenas intangibles de valor —derechos, obligaciones, valores, principios…— dependen estructuralmente de la precondición de disponer de modelos lingüísticos desarrollados computacionalmente y de haber definido también las tácticas de la estrategia de desarrollar modelos generativos fundacionales, LLMs y SLMs en las lenguas naturales específicas de cada cultura y sociedad política. Podrían superarse así (i) los límites que los grandes modelos de lenguaje en lengua inglesa han empezado ya a mostrar[159]; (ii) la separación estricta entre modelos simbólicos y subsimbólicos en IA (mediante el diseño modelos neurosimbólicos con datos multimodales).[160] Estos modelos han empezado a elaborarse para representar la inteligencia de enjambre (*swarm intelligence*). Ésta se refiere a una forma de aprendizaje y toma de decisiones colectiva basada en sistemas descentralizados autoorganizados. Ya se ha sugerido la integración de modelos LLMs en simulaciones multiagente mediante la sustitución de los programas codificados de los agentes por indicaciones (*prompts*).[161]

## 10. Conclusiones y trabajo futuro

Al final, he de volver al principio. He argumentado que el auténtico problema que aporta la introducción de la gestión de la información, el conocimiento y la tecnología mediante técnicas de IA generativa y de técnicas mixtas de IA simbólica y subsimbólica en el mercado, la sociedad y la administración es el de la *reconstrucción de un espacio público en común, global y más justo* donde pueda operar y coevolucionar tanto la inteligencia entre humanos y máquinas como la de aquellos que, lamentablemente, por su situación de pobreza o vulnerabilidad, están excluidos del proceso.

Algunas de las tesis más importantes señaladas en el presente artículo son las siguientes:

1. Existen al menos tres estrategias distintas de regulación de la IA —holística o integral, por principios, y fragmentada o sectorial— según la cultura jurídica en consideración (europea, *Commonwealth*, o norteamericana). El concepto de cadena de valor es utilizado en todas ellas, pero ha sido principalmente integrado en el enfoque holístico de la legislación y política económica europea.
2. Los términos de cadena (genérica) de valor (*value chain*), cadena de valor de suministros (*value supply chain*), cadena global de valor (*global value chain*), y cadena de valor de la IA (*AI value chain, CVIA*) tienen significados distintos que deben ser distinguidos.

---

[158] Cf. Ministerio de Transformación Tecnológica, *ibid*. p. 29.
[159] Cf. Zevallos *et al.* (2023).
[160] Véase *supra*, nota 58 del presente artículo.
[161] Cf. Jimenez-Romero, Yegenoglu, y Blum (2025).

Recientemente, han aparecido también aplicaciones de la CVIA a ámbitos más específicos de la IA —así, la cadena de valor de la infraestructura de la IA (*AI infrastructure value chain*) y la cadena de valor de la IA de uso general mencionada en la Ley europea de la IA (*general-purpose AI value chain*).

3. Aunque es común referirse a la CVIA en el sentido económico de las condiciones y supuestos necesarios para la generación de valor marginal desde el punto de vista monetario, en nuestra acepción es posible referir la CVIA a un marco estructural más amplio para convertir los datos en inteligencia y valor añadido tanto en la dimensión del mercado, como en la social, cultural, ética y jurídica.
4. Desde esta perspectiva normativa, la CVIA se refiere a la creación de valores imbuidos en los sistemas de IA y a la concatenación de derechos y obligaciones en el doble aspecto de (i) la delegación de autonomía, (ii) la agencia moral, de los sistemas de IA. Resulta especialmente relevante la responsabilidad extracontractual, objetiva o condicionada, de la actuación de los agentes.
5. Las disposiciones normativas de la UE —desde los Reglamentos de Protección de Datos, Servicios Digitales, creación del Mercado Europeo y de IA (2020-2024) a las políticas contenidas en la Brújula, el Continente, las Guías Estratégicas, y la ley Ómnibus de 2025, identifican dos pilares conjuntos y complementarios de regulación, uno referido a la implementación de derechos y el segundo a la generación del mercado digital europeo.
6. Resulta sorprendente, por lo tanto, que la reciente política de innovación y competitividad de la IA haya abandonado la vía de la regulación compleja de la CVIA en sus aspectos más intangibles (éticos y jurídicos, especialmente, comprendiendo los derechos humanos), para centrarse en la dimensión estrictamente económica de competitividad y crecimiento, sin atender a la relación que ésta tiene con el desarrollo sustantivo de los derechos, la protección de las garantías, y la necesidad de dar una solución armonizada a los problemas de minimización de riesgos, compensación en caso de daño causado o sobrevenido, y responsabilidad jurídica extracontractual.
7. Este artículo sugiere que una solución integrada a estas cuestiones requiere centrarse nuevamente en la conceptualización de las categorías éticas y jurídicas pertinentes, situando en el centro de la CVIA la propia idea de *validez normativa* que permitiría redefinir qué es lo que entendemos por ecosistemas éticos y jurídicos en la implementación del derecho en los contextos *híbridos* digitales entre humanos y máquinas.

Podemos formular también algunas guías para dirigir este trabajo en el futuro.

Aparte del cálculo genérico de la CV de la IA para superar la disrupción que ha supuesto la introducción de la IA en todas las operaciones de las empresas hasta llegar a la fidelización y comunicación con el consumidor, resulta necesario un desarrollo más preciso de las cadenas normativas de valor éticas y jurídicas, incluyendo sus componentes, los lenguajes para representar su articulación (lógicos y computacionales), las teorías que van a permitir comprender su funcionamiento, y la relación que hay que armar tanto con la cadena económica de valor y la cadena de suministros, como con las cadenas sociales más intangibles (lingüística y cultural).

Habida cuenta de la relevancia del concepto de CVIA en los documentos europeos, sorprende un poco la poca atención que se le ha prestado hasta el momento en los textos de carácter normativo. El trabajo del grupo de expertos en Ética e IA hace referencia a la creación de valor y a la cadena de suministros de una manera genérica. Las preguntas para la autoevaluación de cumplimiento de los siete principios éticos para la IA no incluyen la necesidad de definir y calcular de manera más precisa la minimización de riesgos y la propia CVIA. El Informe sobre la monitorización de la legislación europea mediante IA que establece en realidad el estado del arte en el ámbito de IA y derecho no presta tampoco atención al impacto e implementación de los sistemas de IA en los circuitos de producción y distribución de mercancías y servicios. Si comparamos la reacción de las empresas de datos en enero y febrero de 2025 a la aparición de Deep Seek [162] —ya señalada en la nota 36— veremos la diferencia entre la atención que las corporaciones y empresas prestan a la cadena de valor y la que le han dedicado los expertos en ética y derecho hasta el momento.[163] Más que una diferencia de contenido se trata en realidad de una diferencia de apreciación y de distinta posición en la práctica. El cálculo de la CVIA es necesario para comprender a qué se deben las pérdidas y ganancias en las operaciones de mercado. En mi opinión, también lo es para comprender el grado de implementación de derechos y el tipo de cumplimiento normativo (*compliance*) que se logra en los contextos sociales. Pero esta segunda opción no ha sido explorada con el mismo detenimiento.

La CVIA debe definirse en estrecha relación con (i) la observancia o cumplimiento de la legislación y los principios que los modelos normativos de inteligencia artificial (*compliance*), (ii) la minimización de riesgos, (iii) la delegación de decisiones, (iv) la responsabilidad extracontractual (*non-contractual liability*) y objetiva (*strict liability*) en las situaciones de daño o efectos no deseados producidos por los sistemas (sean multiagente o de IA generativa). Hasta ahora, estos componentes han tendido a considerarse por separado, en abstracto y al margen de su impacto en la sociedad civil, los consumidores, el mercado y las relaciones B2B, y fuera también de la perspectiva holística que permitiría una visión de conjunto. Quizás es hora de darse cuenta de que la CVIA se halla en el centro de la noción de validez jurídica en el momento de implementar los modelos construidos mediante IA en sistemas regulatorios; es decir, en el foco de validez de los ecosistemas éticos y jurídicos basados en diseños normativos.

## 11. Agradecimientos

## 12. Referencias: Bibliografía

---

[162] Cf. Wegner (2025). Véase la explicación en *supra*, nota 36.
[163] El ciclo desde la producción a la comunicación con el consumidor en el mercado minorista (*AI retail chain*) fue planteado y descrito antes de la discusión del Reglamento. Cf. Oosthuizen et al. (2021).

## 13. Referencias: Documentos (China, Canada, España, EU, Francia, ONU, UK, USA)

Pompeu Casanovas

Arequipa, Perú, 16 de octubre de 2025; IIIA-CSIC, Bellaterra, Barcelona, 21 de diciembre de 2025.